\documentclass{SciPost}

\hypersetup{
    colorlinks,
    linkcolor={red!50!black},
    citecolor={blue!50!black},
    urlcolor={blue!80!black}
}

\usepackage[bitstream-charter]{mathdesign}
\DeclareSymbolFont{usualmathcal}{OMS}{cmsy}{m}{n}
\DeclareSymbolFontAlphabet{\mathcal}{usualmathcal}

\fancypagestyle{SPstyle}{
\fancyhf{}
\lhead{\colorbox{scipostblue}{\bf \color{white} ~SciPost Physics }}
\rhead{{\bf \color{scipostdeepblue} ~Submission }}

\fancyfoot[C]{\textbf{\thepage}}
}

\usepackage[mathscr]{eucal}
\usepackage{epsfig,amsfonts}
\usepackage{amsmath}
\usepackage{amsfonts}
\usepackage{amsthm}
\usepackage{graphicx}
\usepackage{hhline}
\usepackage{cite}
\usepackage{dsfont}
\usepackage{psfrag}
\usepackage{mathrsfs} 
\usepackage{hyperref}
\usepackage{array}
\usepackage{tikz}
\usepackage{multirow}
\usepackage{textcomp}
\usepackage{enumitem}
\DeclareMathOperator*{\slim}{slim}
\DeclareMathAlphabet{\mathpzc}{OT1}{pzc}{m}{it}

\makeatletter
\@addtoreset{equation}{section}
\makeatother

\def\be{\begin{equation}}
\def\ee{\end{equation}}
\def\bea{\begin{eqnarray}}
\def\eea{\end{eqnarray}}

\def\ri{{\rm i}}

\newcommand{\rd}{\mbox{d}}
\newcommand{\re}{\mbox{e}}

\newcommand{\Arrw}[4]{
    \draw[thick] (#1-0.05,#2)--(#1-0.05,#2+0.97);
    \draw[thick] (#1+0.05,#2)--(#1+0.05,#2+0.97);
    \draw[thick,out=+160,in=-70]    (#1+0.15,#2+0.9) to (#1,#2+1+0.1);
    \draw[thick,out=+20,in=-110]    (#1-0.15,#2+0.9) to (#1,#2+1+0.1);
    \node[anchor=north] at (#1,#2) {\footnotesize#3};
    \node[anchor=south] at (#1+0.15,#2+0.9) {\footnotesize #4};
}

\begin{document}

\pagestyle{SPstyle}

\begin{center}{\Large \textbf{\color{scipostdeepblue}{
%%%%%%%%%% TODO: Write your article's title here
Scaling limit of the staggered six-vertex model with\\ $U_q\big(\mathfrak{sl}(2)\big)$
invariant boundary conditions\\
}}}\end{center}

\begin{center}\textbf{
% Use (full) first name (+ middle name initials) + surname format.
% Separate subsequent authors by a comma, omit comma and use "and" for the last author.
% Mark the corresponding author(s) with a superscript symbol in this order
% \star, \dagger, \ddagger, \circ, \S, \P, \parallel, ...
Holger Frahm\textsuperscript{$\star$},
Sascha Gehrmann\textsuperscript{$\dagger$} and
Gleb A.\ Kotousov\textsuperscript{$\ddagger$}
}\end{center}

\begin{center}
Institut f\"ur Theoretische Physik, Leibniz Universität Hannover\\
Appelstra\ss{}e 2, 30167 Hannover, Germany
%%%%%%%%%% END TODO: AFFILIATIONS
% Provide email address of corresponding author(s)
\\[\baselineskip]
$\star$ \href{mailto:email1}{\small frahm@itp.uni-hannover.de}\,,\quad
$\dagger$ \href{mailto:email2}{\small sascha.gehrmann@itp.uni-hannover.de}\,,\\
$\ddagger$ \href{mailto:email3}{\small gleb.kotousov@itp.uni-hannover.de}
\end{center}

\section*{\color{scipostdeepblue}{Abstract}}
{\boldmath\textbf{%
We study the scaling limit of a statistical system, which is a special case of the
integrable inhomogeneous six-vertex model. It possesses 
$U_q\big(\mathfrak{sl}(2)\big)$ invariance due to the choice of open boundary conditions
imposed.
An interesting feature of the lattice theory is that the spectrum of scaling dimensions 
contains a continuous component.
By applying the 
ODE/IQFT correspondence and the method of the Baxter $Q$ operator
the corresponding density of  states is obtained. 
In addition, the partition function appearing in the
scaling limit of the lattice model is computed, which may 
be of interest for the study of
nonrational CFTs in the presence of boundaries. As a side result of the research,
a simple formula for the matrix elements of the $Q$ operator for the general, integrable, inhomogeneous six-vertex model was discovered, that has not yet appeared in the literature. It is valid for a certain one parameter family of diagonal open boundary conditions in the sector with the $z$\,-projection of the total spin operator being equal to zero.
}}

\vspace{\baselineskip}

%%%%%%%%%% BLOCK: Copyright information
% This block will be filled during the proof stage, and finilized just before publication.
% It exists here only as a placeholder, and should not be modified by authors.
\noindent\textcolor{white!90!black}{%
\fbox{\parbox{0.975\linewidth}{%
\textcolor{white!40!black}{\begin{tabular}{lr}%
  \begin{minipage}{0.6\textwidth}%
    {\small Copyright attribution to authors. \newline
    This work is a submission to SciPost Physics. \newline
    License information to appear upon publication. \newline
    Publication information to appear upon publication.}
  \end{minipage} & \begin{minipage}{0.4\textwidth}
    {\small Received Date \newline Accepted Date \newline Published Date}%
  \end{minipage}
\end{tabular}}
}}
}
%%%%%%%%%% BLOCK: Copyright information

% For convenience during refereeing we turn on line numbers:
%\linenumbers

\vspace{10pt}
\noindent\rule{\textwidth}{1pt}
\tableofcontents
\noindent\rule{\textwidth}{1pt}
\vspace{10pt}

% Write your article contents here, starting from first \section.

\section{Introduction}
Onsager's solution of the square lattice Ising model \cite{Onsager} opened an era in the application of
integrable systems to the study of phase transitions.
Exactly solvable lattice models in 2D have been found to exhibit 
an interesting array of  effects and powerful analytic/numerical
techniques are available to explore them. Among such phenomena are, for instance,
exactly marginal deformations \cite{Luther,Kadanoff,Barber}, 
strong/weak coupling dualities \cite{KW},
and the appearance of extended conformal symmetry in the scaling limit \cite{Fateev:1985mm}, which
have collectively inspired and refined
our understanding of (conformal) QFT.
Of special mention is 
the Pott's model, a generalization of the 2D Ising model. In the antiferromagnetic case it
possesses a  phase, where the  ground state degeneracy is macroscopic with its
 logarithm  being proportional to the system size; similar to fractons that have recently attracted attention
(see, e.g., \cite{Nandkishore:2018sel} for a review).
This observation was originally made in the 80's by
Berker and Kadanoff \cite{Kadanoff1}. The universal properties were further studied in   \cite{Saleur1}
 via  the mapping of the critical Potts to the  six-vertex model \cite{BaxterPotts} (see also \cite{Baxter:1982zz})
and the application of the  methods of Yang-Baxter integrability and 2D CFT.
\medskip

In the work \cite{Kadanoff1} the Potts model is taken to be 
 homogeneous and isotropic,
so that the coupling between any two nearest neighbour pairs of `spins' is the same. One can consider
an anisotropic version defined on the square lattice, which has two coupling constants --- 
one associated to the vertical and the other to the horizontal edges of the lattice 
that join the  neighbouring spins together. As explained in ref.\,\cite{IJS2}, focusing on
the curve in the parameter space where the model is critical 
leads one to the so-called staggered six-vertex model, which is a special case of the
integrable, inhomogeneous six-vertex model introduced by Baxter in \cite{Baxter:1971}. It gets its name from the fact that
the inhomogeneities are distributed along the square lattice in a checkerboard  (staggered) pattern. In addition,
in the case of the antiferromagnetic Potts model, they 
are fixed to a special value for which the system is `self-dual', i.e., possesses an extra ${\cal Z}_2$ symmetry. 
\medskip

The critical behaviour of the 
staggered six-vertex model at the self-dual point was considered in the works \cite{Jacobsen:2005xz,IJS2}. 
Valuable results about the spectrum of scaling dimensions were obtained by studying the  low energy spectrum 
of the Hamiltonian, which is 
expressed in terms of a logarithmic derivative of the two row transfer-matrix. The Hamiltonian,
unlike the transfer-matrix, is
given by a sum of operators, which act locally on 
$\mathscr{V}_{2L}=\mathbb{C}^2_1\otimes \mathbb{C}^2_2\otimes \ldots\otimes \mathbb{C}^2_{2L}$, where $2L$ is the number
of lattice columns. The precise formula reads as $\mathbb{H}=-\frac{\ri}{q^2-q^{-2}}\sum_{J=1}^{2L}{\cal O}_J$ with\footnote{%
The formula for the Hamiltonian (7.6) in the work  \cite{Bazhanov:2019xvyA} is identical to the one given above except that
the overall sign in front of the  term  $\propto (q-q^{-1})$ in the second line 
of eq.\,\eqref{askj8923qqq}, containing the product of three Pauli matrices, is flipped. This comes about because
that paper uses the different convention for the quantum space:
$\mathscr{V}_{2L}=\mathbb{C}^2_{2L}\otimes \mathbb{C}^2_{2L-1}\otimes \ldots\otimes\mathbb{C}^2_1$, 
see eq.(2.1) therein. The two Hamiltonians are related via the similarity transformation  ${\cal U}\,:\ {\cal U}^2=1$, which
 acts on the local spin operators as ${\cal U}\,\sigma^A_J\,{\cal U}=\sigma_{2L-J+1}^A$.}
\bea\label{askj8923qqq}
{\cal O}_J&=& \big(q-q^{-1}\big)^2\,\sigma^z_J\,\sigma^z_{J+1}+
2\,\big(\sigma^x_J\,\sigma^x_{J+2}+\sigma^y_J\,\sigma^y_{J+2}+
\sigma^z_J\,\sigma^z_{J+2}\big)
\\[0.2cm]
&-&(q-q^{-1})\,\Big(\sigma^z_{J}\,\big(\sigma_{J+1}^x\sigma_{J+2}^x+
\sigma_{J+1}^y\sigma_{J+2}^y\big)
-\big(\sigma_J^x\sigma_{J+1}^x+
\sigma_J^y\sigma_{J+1}^y\big)\,\sigma^z_{J+2}\Big)
-\big(q^2+q^{-2}\big)\, \hat{{\bf 1}}\,.\nonumber
\eea
Here  $\sigma^A_J$ with $A=x,y,z$
are the Pauli matrices that act on site $J$ subject to periodic boundary conditions 
$\sigma_{J+2L}^A=\sigma^A_{J}$, while
the  parameter $q$ is  known as the anisotropy.
The system turns out to be critical when $q$ is a unimodular number and different universal behaviour occurs
depending on   whether $\arg(q)\in(0,\frac{\pi}{2})$ or $\arg(q)\in(\frac{\pi}{2},\pi)$. The regime
\be\label{aksj8923j}
|q|=1\qquad {\rm and}\qquad \arg(q)\in(0,\tfrac{\pi}{2})
\ee
has attracted the most amount of attention. The reason for this is that
 the corresponding spectrum of scaling dimensions was found to possess a continuous component \cite{Jacobsen:2005xz}.
\medskip

The subsequent study of the
regime  \eqref{aksj8923j} saw a remarkable
interchange of ideas between statistical mechanics and formal high energy theory. On the one hand,
the conjecture from ref.\cite{Ikhlef:2011ay},
that the scaling limit of the lattice system is governed by the 2D Euclidean black hole
sigma model introduced in refs.\,\cite{Elitzur:1991cb,Mandal:1991tz,Witten:1991yr}, uses the results of  \cite{Maldacena:2000kv} and its development 
\cite{Hanany:2002ev} which come from the string theory literature. On the other, the detailed study of the vertex model
performed in \cite{Bazhanov:2019xvyA} led to the solution of the spectral problem for the 2D Euclidean black hole CFT, including
the computation of the density of states of the continuous spectrum. 
Perhaps the most surprising output of the research is the following.
While it has been confirmed that {one half} of the partition function arising in the scaling limit
of the vertex model with (quasi-)periodic boundary conditions coincides with the partition function of the 
2D Euclidean black hole sigma model on the torus, the original conjecture of \cite{Ikhlef:2011ay} has been refined. It was proposed in \cite{Bazhanov:2019xvyA} 
that a part of the Hilbert space of the lattice model in the scaling limit should coincide with the pseudo-Hilbert space of the 
black hole sigma model with Lorentzian signature.
\medskip

The above mentioned works all focus on the case
when the lattice is (quasi)-periodic in the horizontal direction. 
In the recent papers \cite{Robertson:2020imc,Frahm:2021ohj}, motivated by the possibility of
making precise contact 
with $D$-brane constructions of non-compact boundary CFTs \cite{Ribault:2003ss,Schomerus:2005aq},
the statistical system has been considered 
with certain integrable, 
open boundary conditions imposed.  In this case the Hamiltonian
is given by
\begin{align}\label{askj8923}
\mathbb{H}\!\!&=&\!\!-\frac{\ri}{q^2-q^{-2}}\bigg(\sum_{J=1}^{2L-2}{\cal O}_J-(q+q^{-1})\,
\big(\sigma_1^x\sigma_2^x+\sigma_1^y\sigma_2^y+
\sigma_{2L-1}^x\sigma_{2L}^x+\sigma_{2L-1}^y\sigma_{2L}^y\big)\nonumber\\[0.2cm]
\!\!&&\!\!
-(q^2-q^{-2})\,(\sigma^z_{2L}-\sigma^z_{1})-2\,(\sigma^z_1\sigma^z_2-\hat{{\bf 1}})+
(q^2+q^{-2})\ (\sigma^z_{2L-1}\sigma^z_{2L}-\hat{{\bf 1}})\bigg)\ ,
\end{align}
where  ${\cal O}_J$ is defined in  eq.\,\eqref{askj8923qqq}. 
A special feature of such a choice of boundary terms is that the model possesses $U_q\big(\mathfrak{sl}(2)\big)$ symmetry.
To explain, 
notice that $\mathbb{H}$ commutes with the $z$\,-projection of the total spin operator:
\be\label{askjj120aaaaa}
\mathbb{S}^z=\frac{1}{2}\,\sum_{J=1}^{2L}\sigma^z_J\, .
\ee
One may check that   it also commutes with
\be\label{askjj12aaaaa}
\mathbb{S}_q^\pm=\ (\mp \ri)\,\sum_{J=1}^{2L}
\bigg(\,\prod_{\ell=J+1}^{2L}q^{-\frac{\sigma^z_\ell}{2}}\,\bigg)\,(-1)^{J}\  \sigma_J^\pm\ 
\bigg(\,\prod_{\ell=1}^{J-1}q^{+\frac{\sigma^z_\ell}{2}}\,\bigg)\ ,
\ee
which, together with $\mathbb{S}^z$, satisfy the defining relations of the $U_q\big(\mathfrak{sl}(2)\big)$ algebra:\footnote{%
The factor $(\mp\ri)$ in 
eq.\,\eqref{askjj12aaaaa}
together with the term $(-1)^{J}$  appearing in the summand  may be removed via a similarity transformation
by a diagonal matrix.}
\be
\big[\mathbb{S}^z\,,\,\mathbb{S}_q^\pm\big]=\pm\mathbb{S}_q^\pm\,,\qquad\qquad
\big[\mathbb{S}_q^+,\mathbb{S}_q^-\big]=\frac{q^{2\mathbb{S}^z}-q^{-2\mathbb{S}^z}}{q-q^{-1}}\,.
\ee
As a result, the eigenstates of the Hamiltonian form irreps of this  algebra. 
These are labelled by the eigenvalues of the Casimir,
\be
2\mathds{C}=(q+q^{-1})\,[\mathbb{S}^z]_q^2+
\mathbb{S}_q^+\,\mathbb{S}_q^-+\mathbb{S}_q^-\,\mathbb{S}_q^+\ ,
\ee
which are given by $[{\cal S}]_q\,[{\cal S}+1]_q$ with integer ${\cal S}=0,1,2,\ldots ,L$
(here we use the standard notation  $[m]_q=(q^m-q^{-m})/(q-q^{-1})$).
The presence of $U_q\big(\mathfrak{sl}(2)\big)$ symmetry simplifies the diagonalization problem for 
$\mathbb{H}$. In particular, one  can restrict
to the $(2L)!/(L!)^2$ dimensional subspace spanned by the eigenstates of the operator $\mathbb{S}^z$,
whose eigenvalues are zero. 
\medskip

It was observed in refs.\,\cite{Robertson:2020imc,Frahm:2021ohj} that the 
spectrum of scaling dimensions of the  staggered six-vertex model
with $U_q\big(\mathfrak{sl}(2)\big)$ invariant open boundary conditions possesses a continuous component
in the regime \eqref{aksj8923j}. 
However, the corresponding density of states 
had not been obtained. This was among the open problems that inspired 
our research.
\medskip

In this paper we perform a systematic study of the low energy spectrum of the Hamiltonian \eqref{askj8923} at large system size $L\gg1$. It is carried out via a mixture of methods,
including a numerical analysis of the
Bethe Ansatz equations as well as the powerful analytical technique of the ODE/IQFT correspondence. It turns out
that the ODEs describing the scaling limit fall within the class of differential equations considered in 
refs.\cite{Bazhanov:2019xvyA,Bazhanov:2019xvy}, where
the universal behaviour of the vertex model with quasi-periodic boundary conditions imposed was studied.
They are described in detail in sec.\,\ref{sec32} below.
As for the Bethe Ansatz equations, they read as
\be\label{akiu32jka}
\bigg(\frac{1+\zeta_m^2q^{+2}}{%
1+\zeta_m^2q^{-2}}\bigg)^{2L}=
q^{4+4{\cal S}}\prod_{j=1\atop j\ne m}^{L-{\cal S}}\frac{\big(\zeta_j-q^{+2}\zeta_m\big)\,
\big(1-q^{+2}\zeta_m\zeta_j\big)}{\big(\zeta_j-q^{-2}\zeta_m\big)\,
\big(1-q^{-2}\zeta_m\zeta_j\big)}\  ,
\ee
where ${\cal S}$ stands for the $U_q\big(\mathfrak{sl}(2)\big)$ total spin of the state.
Having at hand a solution set $\{\zeta_m\}^{L-\mathcal{S}}_{m=1}$ to the above algebraic system, the energy of the corresponding state is 
computed via the formula:
\be\label{anms12i78}
{\cal E}=\sum_{m=1}^{L-{\cal S}}\frac{4\ri\,(q^2-q^{-2})}{\zeta^2_m+\zeta^{-2}_m+q^2+q^{-2}}\ . 
\ee
Notice that eqs.\,\eqref{akiu32jka} are invariant upon making the transformation $\zeta_m\mapsto \zeta_m^{-1}$ 
of any one of the Bethe roots. This allows one, without loss of generality, to assume that
\be
|\zeta_m|\le 1\qquad\qquad (m=1,2,\ldots,L-{\cal S})\, .
\ee
\medskip

The paper is organized as follows. In section \ref{sec2} we present a formula for the matrix elements of the
Baxter $Q$ operator that was used in our numerical work. It is
valid in the sector $S^z=0$ and for a one parameter family of open boundary conditions.
The outcomes of our study of the low energy spectrum of the Hamiltonian \eqref{askj8923} 
in the regime \eqref{aksj8923j} is given in
section \ref{sec3}. The first subsection thereof, for the most part, is a review of the results contained in 
refs.\cite{Robertson:2020imc,Frahm:2021ohj}. In the next two, the ODE/IQFT correspondence is
described and, on the basis of this, the so-called `quantization condition' is obtained. The latter 
is what allows us to perform a full characterization of the low energy space of states of the lattice model, 
which is detailed in section \ref{sec4}. In section \ref{sec5}, the formula
for the partition function appearing in the scaling limit of the lattice model is given, which may be of interest
to those studying non-rational boundary CFT. The last section is devoted to a discussion and includes
a summary of the main results.

\section{The \texorpdfstring{$Q$}{Q} operator\label{sec2}}
In analysing the low energy spectrum of the Hamiltonian of a critical lattice system in the scaling limit
one meets with immediate issues. An important one concerns
the construction of RG trajectories, where low energy stationary states at different lattice sizes $L$
are grouped into families $\{|\Psi_L\rangle\}$. It is clear how to define the RG flow for the ground
state or, for that matter, the lowest energy states in the disjoint sectors of the Hilbert
space, say, the sector with given eigenvalue of $\mathbb{S}^z$ \eqref{askjj120aaaaa}. 
However, assigning an $L$ dependence to
 a   low energy excited state  and then continuing $|\Psi_L\rangle$ 
to $L\gg 1$ seems to be a non-trivial task. 
\medskip

 In the case at hand the model is integrable and
the construction of RG trajectories 
is facilitated by the  Bethe Ansatz solution. A formulation of the procedure is provided in ref.\,\cite{Bazhanov:2019xvy},
see also the work \cite{Bazhanov:2019xvyA}.
The low energy stationary states can be chosen to be the Bethe states,
which are simultaneous eigenvectors of the full family of  operators commuting with the Hamiltonian. 
They are labeled by the Bethe roots $\{\zeta_m\}$,
which solve the Bethe Ansatz equations \eqref{akiu32jka}. The 
eigenvalues of the commuting operators for the Bethe state are given in terms of
the corresponding  $\{\zeta_m\}$,
as in formula \eqref{anms12i78} for the energy, and their computation does
not require any explicit diagonalization. Suppose one has at hand the Bethe roots
for a low energy Bethe state for a lattice of $L=L_{\rm in}$ sites. The state
$|\Psi_{L_{\rm in}+2}\rangle$ is specified such that the pattern of Bethe roots qualitatively
remains the same. These can be obtained by numerically solving the Bethe Ansatz equations 
\eqref{akiu32jka}, where the initial approximation for the iterative solution
 is constructed from the Bethe roots corresponding to $|\Psi_{L_{\rm in}}\rangle$.
By iterating this procedure an RG trajectory $\{|\Psi_L\rangle\}$ for
 increasing $L$ is obtained.
\medskip

\begin{figure}[t]
    \centering
    
    \begin{minipage}[b]{.49\linewidth}
    \scalebox{0.85}{
\begin{tikzpicture}
\node at (3,1) { $u$};
\draw (3,1.01) circle (0.25cm);
\node at (0,0) {\includegraphics[width=0.9\linewidth]{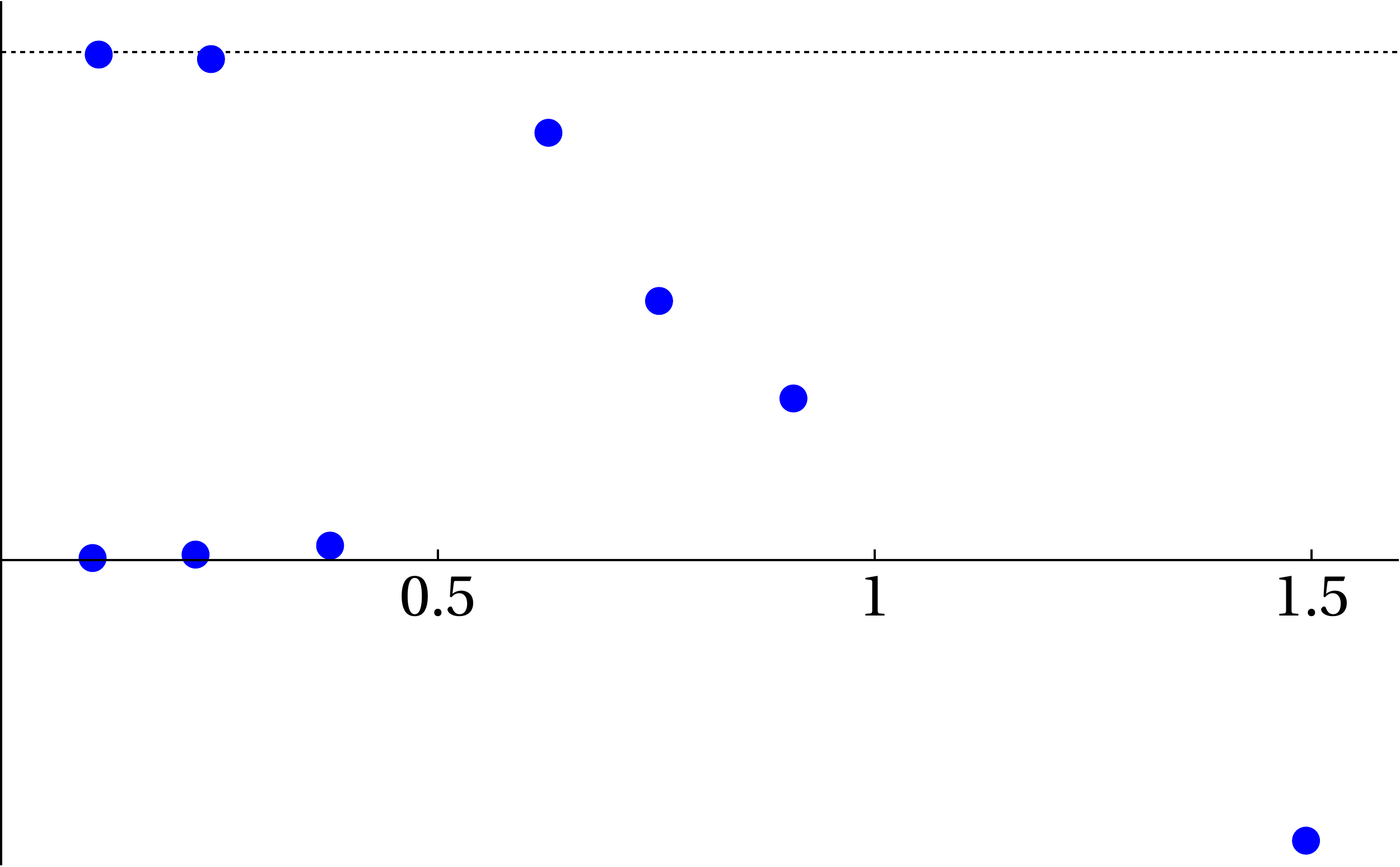}};
\node[anchor=west] at (1.75,2.35)   {$\Im m(u)=\frac{\pi}{2}$};
\end{tikzpicture}
}
 \end{minipage}
 \begin{minipage}[b]{.49\linewidth}
    \scalebox{0.85}{
\begin{tikzpicture}
\node at (3,1) { $u$};
\draw (3,1.01) circle (0.25cm);
\node at (0,0) {\includegraphics[width=0.9\linewidth]{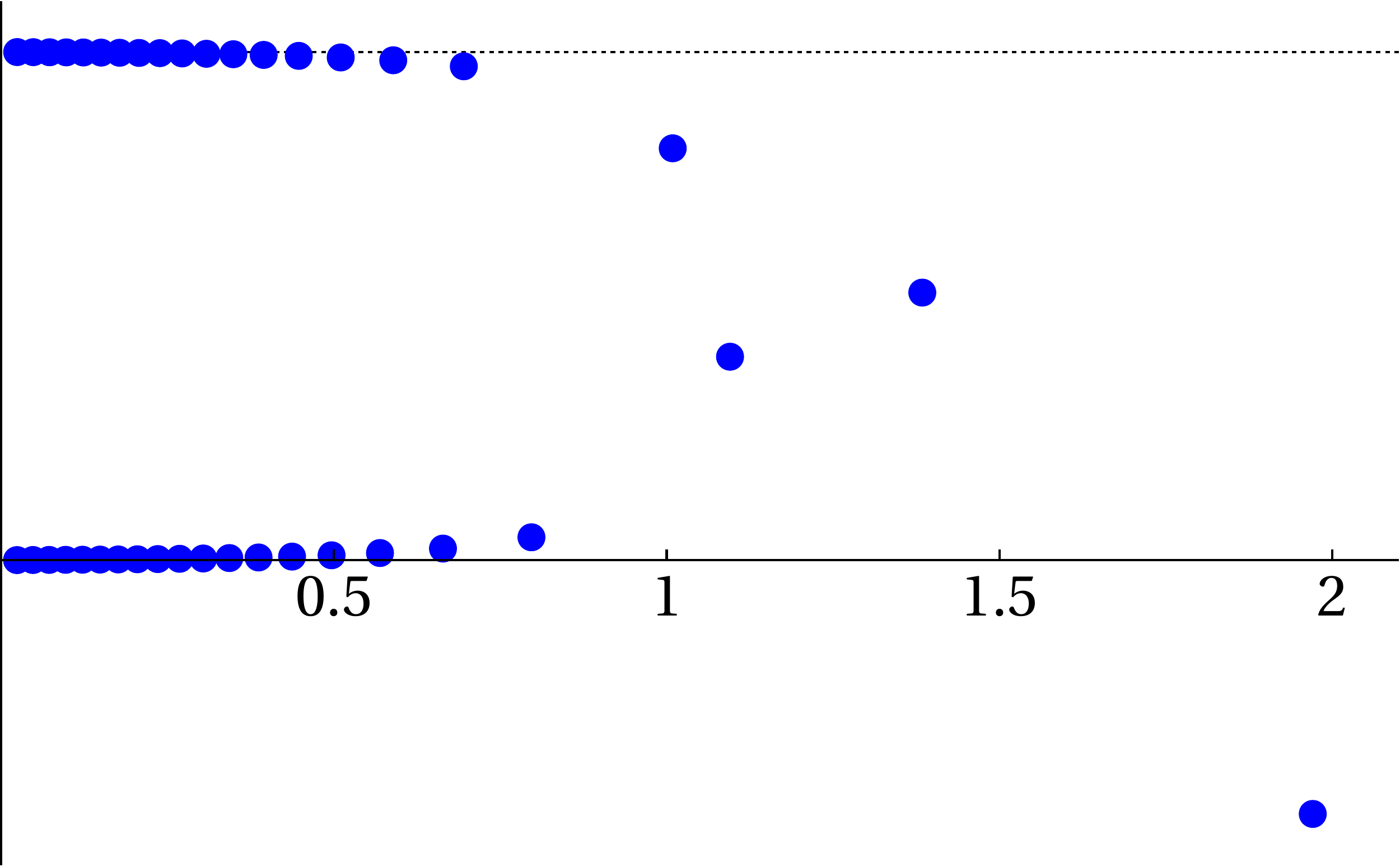}};
\node[anchor=west] at (1.6,2.35)   {$\Im m(u)=\frac{\pi}{2}$};
\end{tikzpicture}
}
 \end{minipage}

    \caption{\small
Displayed is the pattern of Bethe roots in the complex $u=-\frac{1}{2}\log(\zeta)$ plane for a low energy Bethe state
that was generically chosen. For the left panel $L=10$ and the set $\{\zeta_m\}$ was obtained from the eigenvalue of the $Q$ operator computed on the state.  The latter was used to seed an RG trajectory $\{|\Psi_L \rangle\}$ and, on the right panel, shown are the Bethe roots for the corresponding state with $L=40$. The trajectory is labelled by $\mathcal{S}=1$, while the
anisotropy parameter was taken to be $q=\re^{\frac{10 \pi\ri }{49}}$. \label{Generic_State}}
    % State number is 53
    % Saved under spectro/RG/Z_53
\end{figure}
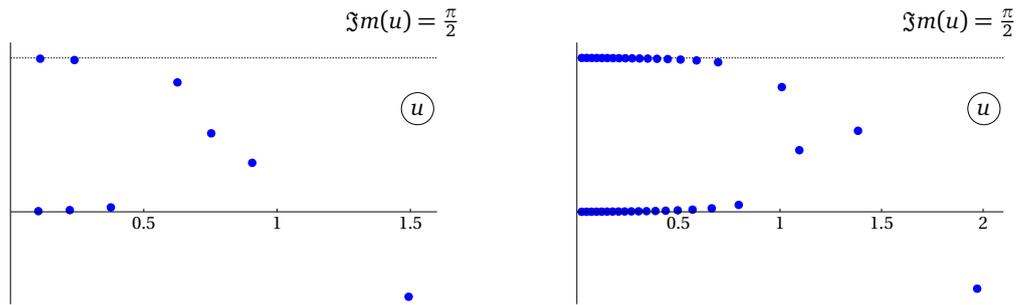

A method is required in order to extract the Bethe roots   corresponding to the state $|\Psi_{L_{\rm in}}\rangle$
which seeds the RG trajectory. Typically for $|\Psi_{L_{\rm in}}\rangle$ generic, $\{\zeta_m\}$ are complex numbers that
do not resemble a simple pattern in the complex plane, see Fig.\,\ref{Generic_State} for an example.
Finding all of the possible 
solution sets of the Bethe Ansatz equations  \eqref{akiu32jka} 
even  for $L_{\rm in}\lesssim 10$ is impossible to carry out on a modern laptop
because the system is too complicated.
Moreover, searching for the set $\{\zeta_m\}$ by applying the Newton method
to \eqref{akiu32jka}, trying out various
initial approximations, is time consuming and not guaranteed to work.
\medskip

In our study of the spin chain,
to find the Bethe roots corresponding to a generic low energy state $|\Psi_{L_{\rm in}}\rangle$ systematically,
 we used the technique 
based on  the notion of the  Baxter $Q$ operator  $\mathbb{Q}(\zeta)$ \cite{Baxter:1972hz}. The latter commutes with
itself for different values of the spectral parameter as well as the transfer-matrix,
\be\label{Commutativity_Q_T}
\mathbb{Q}(\zeta)\,: \ \qquad [\mathbb{Q}(\zeta),\mathbb{Q}(\zeta')]=[\mathbb{Q}(\zeta),\mathbb{T}(\zeta')]=0\,.
\ee
In addition, its eigenvalues  for a Bethe state is a polynomial in $\zeta$, whose
 zeroes coincide with the corresponding solution set to the Bethe Ansatz equations (see, e.g., formula \eqref{kash12bv32ds} below).
For a small number of lattice sites it is possible to obtain
 the first few hundred low lying stationary states of the spin chain  via an explicit
diagonalization of the Hamiltonian. For each of them,
the corresponding set $\{\zeta_m\}$ is extracted by computing the eigenvalue 
of $\mathbb{Q}(\zeta)$  (some further details
can be found in section 18 of ref.\cite{Bazhanov:2019xvyA}). In turn, the RG trajectory $\{|\Psi_L\rangle\}$ is continued from $L=L_{\rm in}$ to large $L$ via
the procedure outlined above.
\bigskip

Below we present the explicit formula for the $Q$ operator
that was used in our analysis. It allowed us to go beyond the results of the previous
papers \cite{Robertson:2020imc,Frahm:2021ohj}.
Our formula was obtained  based on the results of ref.\cite{Tsuboi:2020uoh}, which is part of an  interesting group of recent papers
\cite{Tsuboi:2020uoh,Frassek:2015mra,Baseilhac:2017hoz,VlaarWeston}
appearing in the mathematics literature. 
Among other things, they give $\mathbb{Q}(\zeta)$ as a trace 
of a monodromy matrix over a $q$-oscillator representation for  models associated with
the rational \cite{Frassek:2015mra} and trigonometric \cite{Baseilhac:2017hoz,VlaarWeston}
 $R$-matrix for $\mathfrak{sl}(2)$  with a two parameter family of open Boundary Conditions (BCs).
The matrix elements of $\mathbb{Q}(\zeta)$ from \cite{Tsuboi:2020uoh}  take the form of an infinite sum,
which converges only in a certain parameteric domain that excludes the model with $U_q\big(\mathfrak{sl}(2)\big)$ 
invariant BCs we are considering. As such,
some analysis was required in order to bring the expression to a form, which is literally applicable to the case at hand
and is efficient for numerical purposes. 
\bigskip

Let's consider the more general case of a lattice system  with an arbitrary set of inhomogeneities
$\{\eta_J\}_{J=1}^{2L}$.
Moreover, for technical reasons, 
we'll take  a one parameter family of open BCs depending on $\epsilon$, 
such that the $U_q\big(\mathfrak{sl}(2)\big)$
invariant case is recovered at
\be
\epsilon=0 \qquad\qquad
\qquad\qquad\Big(U_q\big(\mathfrak{sl}(2)\big)\ {\rm invariant}\ {\rm BCs}\Big)\ .
\ee
Rather than following the conventions of ref.\cite{Frahm:2021ohj},
we'll use the multiplicative spectral parameter $\zeta$ and arrange
the definitions such that
the matrix elements of both $\mathbb{T}(\zeta)$ and 
$\mathbb{Q}(\zeta)$ are mainifestly polynomials in $\zeta$.
The $R$-matrix reads as
\be\label{Rmat1a}
R(\zeta)=\left(\begin{array}{cccc}
q-q^{-1}\,\zeta& 0 & 0 & 0 \\[0.0cm]
0 & 1-\zeta & q-q^{-1}& 0 \\[0.0cm]
0 & (q-q^{-1})\,\zeta& 1-\zeta & 0 \\[0.0cm]
0 & 0 & 0 & q-q^{-1}\,\zeta
\end{array}\right)\,. 
\ee
It can be interpreted as an operator $R(\zeta)=R_{I,J}(\zeta)$ acting on the tensor product $\mathbb{C}_I^2 \otimes \mathbb{C}^2_{J}$, where 
 $\mathbb{C}_J^2$ with $J=0$ stands for the auxiliary space, while for $J=1,\dots ,2L$ it is the $J$-th factor of the quantum space
\be\label{askb321}
\mathscr{V}_{2L}=\mathbb{C}_{1}^2\otimes \mathbb{C}_{2}^2\otimes\ldots\otimes \mathbb{C}_{2L}^2\ .
\ee
\begin{figure}[t]
\centering
\scalebox{1.0}{
\begin{tikzpicture}
\node[anchor=west] at (1.5,-0.075) {$\mathbb{T}(\zeta q^{-1})=$};

\node[anchor=west] at (4,-0.125) {$K^+$};
\node[anchor=west] at (11.25,-0.125) {$K^-$};
\draw[thick,-<] (5.2,0)--(8.1,0);
\draw[thick,-] (8,0)--(10.8,0);

\draw[thick,-] (10.8,0)--(11,-0.1);
\draw[thick,-] (11,-0.1)--(10.8,-0.2);

\draw[thick,-] (5.2,0)--(5,-0.1);
\draw[thick,-] (5,-0.1)--(5.2,-0.2);

\draw[thick,->] (5.2,-0.20)--(8.1,-0.20);
\draw[thick,-] (8.0,-0.20)--(10.8,-0.20);

\Arrw{0.4+8}{-0.5}{$\eta_{4}$}{$\frac{1}{2}$}
\Arrw{1.2+8}{-0.5}{$\eta_{3}$}{$\frac{1}{2}$}
\Arrw{2.+8}{-0.5}{$\eta_2$}{$\frac{1}{2}$}
\Arrw{2.75+8}{-0.5}{$\eta_1$}{$\frac{1}{2}$}
\filldraw (-0.2+7.2,-0.75) circle (0.3pt);
\filldraw (0+7.2,-0.75) circle (0.3pt);
\filldraw (0.2+7.2,-0.75) circle (0.3pt);

\filldraw (-0.2+7.2,0.7) circle (0.3pt);
\filldraw (0+7.2,0.7) circle (0.3pt);
\filldraw (0.2+7.2,0.7) circle (0.3pt);

\Arrw{-0.4+8}{-0.5}{}{}
\Arrw{-1.2+8}{-0.5}{}{}
\Arrw{-2.0+8}{-0.5}{$\eta_{2L-1}$}{$\frac{1}{2}$}
\Arrw{-2.75+8}{-0.5}{$\eta_{2L}$}{$\frac{1}{2}$}
\end{tikzpicture}}

    \caption{\small
 A graphical representation of the transfer-matrix for the inhomogeneous six-vertex model. Open
BCs are imposed, as indicated by the presence of the reflection matrices $K^\pm$ given by (\ref{K_Matrices}) acting  in the two-dimensional auxiliary space.}
    \label{fig:Transfer}
\end{figure}
The transfer-matrix, which is graphically depicted in Fig.\,\ref{fig:Transfer},  is given by
\be\label{Transfer_matrix_Diagonal_BC}
    \mathbb{T}(\zeta q^{-1})=q^{-2L}\, \mathrm{tr} _0 \Big( K^+_0(\zeta )\, R_{0,2L}(\zeta\,
  \eta^{-1}_{2L})\dots R_{0,1}(\zeta \, \eta^{-1}_1)\,
    K^{-}_0(\zeta )\, R_{1,0}(\zeta\,  \eta_1)\dots R_{2L,0}(\zeta\,  \eta_{2L})  \Big)\,.
\ee
Here the trace is taken over the auxiliary space, while $K^\pm$ stand for the diagonal matrices 
\be\label{K_Matrices}
K^-(\zeta)=\begin{pmatrix} 1 +\zeta\epsilon&0\\
    0& \zeta^{2} + \zeta\epsilon\end{pmatrix},\quad \qquad     K^+(\zeta)=\begin{pmatrix} q^{-2}\zeta + \epsilon&0\\
    0& \zeta^{-1} +q^{-2}\epsilon\end{pmatrix}\,.
\ee
It is straightforward to check that $\mathbb{T}(\zeta)$
 is a polynomial of order $4L+2$ in the spectral parameter and satisfies the conditions
\be\label{askj1287ad}
\mathbb{T}(0)=\epsilon\, \big(q^{2\mathbb{S}^z}+q^{-2\mathbb{S}^z}\big)\ ,\qquad\qquad\qquad
\mathbb{T}(\zeta^{-1})=\zeta^{-4L-2}\ \mathbb{T}(\zeta)\ .
\ee
To get back the case of the staggered six-vertex model with $U_q\big(\mathfrak{sl}(2)\big)$ invariant BCs, at the self-dual point, 
one should fix the inhomogeneities $\eta_J$ and the extra parameter $\epsilon$ to be
\be\label{mansiu12hj}
\eta_{J}=\ri\,(-1)^{J}\qquad\qquad
(J=1,2,\ldots 2L)\,,\qquad\qquad \epsilon=0\,.
\ee
We will use the notation\footnote{%
The transfer-matrix as defined in formula \eqref{Transfer_matrix_Diagonal_BC} for arbitrary inhomogenieties  $\eta_J$ and $\epsilon=0$
is related to $\tau(u)$ given by (2.9) of ref. \cite{Frahm:2021ohj} via a similarity transformation by a diagonal matrix and an overall 
multiplicative factor. Namely, \begin{equation*} \mathbb{T}^{(0)}(\zeta)= 2^{4L} \re^{-4 L u-2 u} \bigg(\prod^{2L}_{J=1} \rho(-u+\tfrac{\ri \gamma}{2}-\delta_J) \bigg)\ \mathcal{U}\, \tau(u)\,\mathcal{U}^{-1}. \end{equation*}  Here $\mathcal{U}=G_1(\delta_1)\otimes \dots \otimes G_{2L}(\delta_{2L})$ with $G(u)={\rm{diag}}(1,\re^{-u}) $,  while the parameters $u,\gamma,\delta_J$ need to be identified with $\zeta,q,\eta_J$ as
\begin{equation*}  \zeta=\re^{-2u}, \qquad q=\re^{\ri \gamma}, \qquad \eta_J=\re^{-2\delta_J} \,.\end{equation*}
Also, the function $\rho(u)$ is given by $\rho(u)=\frac{1}{2}\big(\cos(2\gamma)-\cosh(2u)\big)$.
}
\be
\mathbb{T}^{(0)}(\zeta)\equiv \mathbb{T}(\zeta)\big|_{\epsilon=0}\,.
\ee
\bigskip

For our purposes it is sufficient to focus on the sector where the eigenvalue of the $z$\,-projection of the total
spin operator is zero, i.e., $S^z=0$. This is because for the model possessing  $U_q\big(\mathfrak{sl}(2)\big)$ invariance, the states come in multiplets $\mathcal{M}_{\mathcal{S}}$ each of which has a representative in that sector. Let the tuples $(a_1a_2\dots a_{2L})$ and $(b_1b_2\dots b_{2L})$  with $a_J,b_J=\pm$ be the input/output indices for the space $\mathscr{V}_{2L}$ \eqref{askb321}. To present the formula for the  $Q$ operator, introduce
\bea\label{aksj81}
&&\left(\begin{array}{cc}
\big[{A}(\zeta;m)\big]^{+}_{+} & \big[{A}(\zeta;m)\big]^{-}_{+} \\[0.2cm]
\big[{A}(\zeta;m)\big]^{+}_{-} & \big[{A}(\zeta;m)\big]^{-}_{-}
\end{array}\right)=\left(\begin{array}{cc}
q^{m} &   q^{m}   \\[0.2cm]
  \zeta q^{-m+1}    & q^{-m}
\end{array}\right)\,,\\[0.25cm]
&&\left(\begin{array}{cc}
\big[{\widetilde{A}}(\zeta;m)\big]^{+}_{+} & \big[{\widetilde{A}}(\zeta;m)\big]^{-}_{+} \\[0.2cm]
\big[{\widetilde{A}}(\zeta;m)\big]^{+}_{-} & \big[{\widetilde{A}}(\zeta;m)\big]^{-}_{-}
\end{array}\right)=\left(\begin{array}{cc}
q^{m} &   \zeta q^{m+2}   \\[0.2cm]
   q^{-m-1}    & q^{-m}
\end{array}\right)\,.
\eea
The matrix elements of $\mathbb{Q}(\zeta)$, valid in the sector $S^z=0$, are given by\footnote{Written in index notation, the transfer matrix \eqref{Transfer_matrix_Diagonal_BC} takes the form:
\bea
   \big[ \mathbb{T}(\zeta q^{-1})\big]^{b_1\dots b_{2L}}_{a_1\dots a_{2L}}=q^{-2L}\sum_{c_1\dots c_{2L}=\pm}\sum_{\alpha_1\dots \alpha_{2L+1}=\pm \atop \beta_1\dots \beta_{2L+1}=\pm } && \big[K^+(\zeta )\big]^{\beta_{2L+1}}_{\alpha_{2L+1}}\, \big[R(\zeta  \eta^{-1}_{2L})\big]^{\alpha_{2L+1}b_{2L}}_{\alpha_{2L}c_{2L}}\dots \big[R(\zeta  \eta^{-1}_1)\big]^{\alpha_2 b_1}_{\alpha_1 c_1}\,\nonumber\\[0.0cm]
    &&\big[K^{-}(\zeta )\big]^{\alpha_1}_{\beta_1}\, \big[R(\zeta  \eta_1)\big]^{c_1\beta_1}_{a_1 \beta_2}\dots \big[R(\zeta  \eta_{2L})\big]^{c_{2L}\beta_{2L}}_{a_{2L}\beta_{2L+1}}  \,.\nonumber
\eea Here $\big[K^{\pm}(\zeta)\big]^{\beta}_{\alpha}$ with $\alpha,\beta=\pm$ stand for the entries of the diagonal matrices \eqref{K_Matrices} and similar for $\big[R(\zeta)\big]^{\beta b}_{\alpha a}$, e.g., $\big[R(\zeta)\big]^{+-}_{-+}=q-q^{-1}$ and $\big[R(\zeta)\big]^{-+}_{+-}=(q-q^{-1})\,\zeta$.
}
\be\label{asnb3bv}
\big[\mathbb{Q}(\zeta)\big]^{b_{1}\dots b_{2L}}_{a_{1}\dots  a_{2L}}= \sum_{c_1\dots c_{2L}=\pm} q^{(S^z_c)^2} (\epsilon q^{2}\zeta)^{S^z_c}  \prod^{2L}_{J=1}\big[{A(\zeta\eta^{-1}_J,m_{b,J})}\big]^{b_J}_{c_J}\, \big[{\widetilde{A}}(\zeta\eta_J,m_{a,J+1})\big]^{c_J}_{a_J}\,.
\ee
Here the symbol $S_{\rm c}^z$, which should not be confused with the eigenvalue of the $z$\,-projection of the total spin operator,  stands for
\be\label{askb3vb12}
S_{\rm c}^z=\frac{1}{2}\sum_{J=1}^{2L} c_J
\ee
and provides a natural grading for the sum over $c_J$. 
The internal indices $\{m_{a,J},m_{b,J}\}$ 
come from the product over the auxiliary space. They
are  fixed completely by the ice-rule to be
\be\label{ajsbvb3vhg}
m_{x,J}=\frac{1}{2}\sum_{\ell=J}^{2L} (x_\ell-c_\ell)
\qquad\qquad (J=1,\ldots,2L+1;\  x=a,b)\, .
\ee
Despite the presence of the factor $\zeta^{S^z_c}$ in formula \eqref{asnb3bv}, where the exponent can  be negative,
 the matrix elements of $\mathbb{Q}(\zeta)$ turn out to be polynomials in $\zeta$ of degree $2L$. Moreover, it is straightforward to show that they satisfy:
\be\label{Properties_Q}
\mathbb{Q}(\zeta^{-1})=\zeta^{-2L}\ \mathbb{Q}(\zeta)\,, \qquad\qquad  
\mathbb{Q}(0)={\bf 1}\,.
\ee
\medskip

The following comment is in order  regarding the relation of the $Q$ operator whose matrix elements are given by
  \eqref{asnb3bv} with that studied  in ref.\,\cite{Tsuboi:2020uoh}. In fact, in  the latter work  two $Q$ operators
$\mathbb{Q}^{(a)}$ with $a=1,2$ are introduced. In order to specialize
to the one parameter family of open boundary conditions being considered here one should restrict the parameters
$\bar{\epsilon}_\pm$ and $\epsilon_\pm$ in that paper as
$\epsilon_+/\epsilon_-=\bar{\epsilon}_+/\bar{\epsilon}_-=\epsilon$. Then, choosing a representation
for the $q$ oscillator algebra, formula  (5.13) from \cite{Tsuboi:2020uoh} provides an expression for
 the matrix elements of the $Q$ operators in terms of an infinite sum $\sum_{m\ge 0} g(m)$ . For the sector $S^z=0$
and generic values of the parameters
this sum diverges for both $\mathbb{Q}^{(1)}$ and $\mathbb{Q}^{(2)}$ so that eq.\,(5.13)
becomes inapplicable. This is because the summand $g(m)$
tends to a finite, nonvanishing limit as $m\to\infty$. Our formula \eqref{asnb3bv}
essentially coincides with $\lim_{m\to\infty}g(m)$. The limiting value is the same whether or not we started from
$\mathbb{Q}^{(1)}$ or $\mathbb{Q}^{(2)}$ so one may only obtain a single $Q$ operator in this way. Remarkably,
we have checked for small lattice sizes $L=2,3,4,\ldots$ that $\mathbb{Q}(\zeta)$ from  \eqref{asnb3bv}
obeys the commutativity conditions \eqref{Commutativity_Q_T}
as well as the  $TQ$ relation with the transfer-matrix
\bea\label{kjas871jk}
\big(1-\zeta^{2}\big)\,\mathbb{Q}(\zeta)\,\mathbb{T}(\zeta)&=&
\,\big(\epsilon+q^{+1}\,\zeta\big)\,\big(1+\zeta\,q^{+1}\,\epsilon\big)\, f(q^{-1}\,\zeta)\ 
\mathbb{Q}\big(\zeta\,q^{+2}\big) \\[0.2cm]
&+&
\big(\epsilon+q^{-1}\,\zeta\big)\,\big(1+\zeta\,q^{-1}\,\epsilon \big)\,
f(q^{+1}\zeta)\ \mathbb{Q}\big(\zeta\,q^{-2}\big)\qquad     \qquad (S^z=0)\,.\nonumber
\eea
 Here
\be\label{askjh21}
f(\zeta)=(1-\zeta^{2})
\prod_{J=1}^{2L}\big(\zeta-\eta_J^{-1}\big)
\big(\zeta-\eta_J\big)\ ,
\ee
while the values of the inhomogeneities  $\eta_J$ and
parameter $\epsilon$ are assumed to be generic.
The Bethe Ansatz equations for the integrable model follow as usual: One considers both sides of eq.\,\eqref{kjas871jk} evaluated on a common eigenvector. In view of \eqref{Properties_Q}, the eigenvalues of 
$\mathbb{Q}(\zeta)$ take the form
\be\label{kash12bv32ds}
Q(\zeta)=\prod_{j=1}^L\,(1-\zeta/\zeta_j)\,(1-\zeta\zeta_j)\,.
\ee
Combining the above with \eqref{kjas871jk} and setting  $\zeta=\zeta_m$ into that formula, one arrives at the coupled system of algebraic equations for the Bethe roots:
\bea\label{akiu32jkaAAA}
\prod_{J=1}^{2L}\frac{\big(\zeta_mq^{+1}-\eta_J^{-1}\big)\,\big(\zeta_mq^{+1}-\eta_J\big)}{%
\big(\zeta_mq^{-1}-\eta_J^{-1}\big)\,\big(\zeta_mq^{-1}-\eta_J\big)}&=&
\frac{\big(\epsilon+q^{+1}\,\zeta_m\big)\,\big(q^{+1}\,\epsilon+\zeta^{-1}_m\big)}{%
\big(\epsilon+q^{-1}\,\zeta_m\big)\,\big(q^{-1}\,\epsilon +\zeta^{-1}_m\big)}
\\[0.2cm]
&\times&
q^2\prod_{j=1\atop j\ne m}^L\frac{\big(\zeta_j-q^{+2}\zeta_m\big)\,
\big(1-q^{+2}\zeta_m\zeta_j\big)}{\big(\zeta_j-q^{-2}\zeta_m\big)\,
\big(1-q^{-2}\zeta_m\zeta_j\big)}\qquad\qquad (S^z=0)\, .\nonumber
\eea
These are equivalent to the Bethe Ansatz equations obtained in the original work \cite{Sklyanin:1988yz},
specialized to the sector $S^z=0$ and with
  $\xi^{+}=- \xi^{-}$. Also, the parameters should be identified as 
$(q,\eta_J,\epsilon)\mapsto (\re^{\eta},\re^{-2u_J},\re^{2\xi^\pm})$, while
 $\zeta_m\mapsto  \re^{-2v_m}$. We find it surprising that a formula like \eqref{asnb3bv} exists 
and its further exploration may be worthwhile. At the same time, since our work is focused on
the study of the scaling limit of an integrable lattice system, we believe that it is not the place to do this here.

\medskip

\medskip

Some care is needed in taking the limit $\epsilon\to 0$ of the $Q$ operator.
It is clear from the explicit formula \eqref{asnb3bv}  that the matrix elements of 
$\mathbb{Q}(\zeta)$ generically diverge due to the presence of the factor $\epsilon^{S^z_{\rm c}}$, where the exponent may be negative.
This is a manifestation of the $U_q\big(\mathfrak{sl}(2)\big)$ invariance possessed by 
the model at the point $\epsilon=0$, so that states in different sectors of $S^z$ form multiplets of
the symmetry group that have the same eigenvalue of $\mathbb{Q}(\zeta)$.
In ref.\cite{Bazhanov:2010ts} a similar phenomenon was studied in the context of the
XXX spin chain with twisted boundary conditions controlled by the parameter $\phi$.
At $\phi=0$ the model possesses global $\mathfrak{su}(2)$ symmetry and the matrix elements
of the $Q$ operator become infinite. 
It was explained in that paper how to  take the limit $\phi\to0$ 
of $\mathbb{Q}(\zeta)$ so that one obtains a well defined result.
The discussion is readily adapted to the case at hand.
\medskip

Recall that the quadratic Casimir for the $U_q\big(\mathfrak{sl}(2)\big)$ algebra is given by:
\be
2\mathds{C}=(q+q^{-1})\,[\mathbb{S}^z]_q^2+
\mathbb{S}_q^+\,\mathbb{S}_q^-+\mathbb{S}_q^-\,\mathbb{S}_q^+\ ,
\ee
where
$\mathbb{S}^z$ is defined in formula \eqref{askjj120aaaaa} in the introduction, while for arbitrary values of the inhomogeneities,
\be
\mathbb{S}_q^\pm=\sum_{J=1}^{2L}
\bigg(\,\prod_{\ell=J+1}^{2L}q^{-\frac{\sigma^z_\ell}{2}}\,\bigg)\ \eta_J^{\mp 1} \sigma_J^\pm\ 
\bigg(\,\prod_{\ell=1}^{J-1}q^{+\frac{\sigma^z_\ell}{2}}\,\bigg)\ .
\ee
The  eigenvalues  of $\mathds{C}$ are given by
\be
[{\cal S}]_q\,[{\cal S}+1]_q
\ee
with   ${\cal S}=0,1,2,\ldots,L\ $.
One can consider ${\cal S}$ as an operator, which for an eigenstate of the quadratic Casimir 
with eigenvalue $[{\cal S}]_q\,[{\cal S}+1]_q$ gives back 
the non-negative integer ${\cal S}\ge 0$.
Then, following the work \cite{Bazhanov:2010ts}, it turns out that 
 the limit 
\be\label{kjsahg3}
\mathbb{Q}^{(0)}(\zeta)=\lim_{\epsilon\to 0}\,\epsilon^{\frac{{\cal S}}{2}}\,\mathbb{Q}(\zeta;\epsilon)\,
\epsilon^{\frac{{\cal S}}{2}}
\ee 
exists and yields 
the $Q$ operator for the inhomogeneous six-vertex model with $U_q\big(\mathfrak{sl}(2)\big)$ invariant
BCs in the sector $S^z=0$. The commutativity condition \eqref{Commutativity_Q_T} and the $TQ$ relation \eqref{kjas871jk} with the substitutions $\big(\mathbb{Q},\mathbb{T}\big)\mapsto 
\big(\mathbb{Q}^{(0)},\mathbb{T}^{(0)}\big)$ are satisfied provided, for the latter formula, one sets $\epsilon=0$. We note, however, that the normalisation \eqref{Properties_Q} no longer holds true. Instead, 
\be\label{Eigenvalue_Q_0_BR}
\mathbb{Q}^{(0)}(\zeta)\,|\Psi_{L}\rangle=C\,\zeta^{\mathcal{S}}
\prod_{j=1}^{L-\mathcal{S}}\,(1-\zeta/\zeta_j)\,(1-\zeta\zeta_j)\,|\Psi_{L}\rangle\,,
\ee
where $C$ is a constant  depending only on $q$ and $\mathcal{S}$. 
\medskip

The $TQ$ relation specialized to $\epsilon=0$, combined with \eqref{Eigenvalue_Q_0_BR}, leads to the Bethe Ansatz equations
\be\label{skja8932}
\prod_{J=1}^{2L}\frac{\big(\zeta_mq^{+1}-\eta_J^{-1}\big)\,\big(\zeta_mq^{+1}-\eta_J\big)}{%
\big(\zeta_mq^{-1}-\eta_J^{-1}\big)\,\big(\zeta_mq^{-1}-\eta_J\big)}=
q^{4+4\mathcal{S}}\prod_{j=1\atop j\ne m}^{L-\mathcal{S}}\frac{\big(\zeta_j-q^{+2}\zeta_m\big)\,
\big(1-q^{+2}\zeta_m\zeta_j\big)}{\big(\zeta_j-q^{-2}\zeta_m\big)\,
\big(1-q^{-2}\zeta_m\zeta_j\big)}\  .
\ee
Upon taking the inhomogeneities to be as in \eqref{mansiu12hj} one gets back
eq.\,\eqref{akiu32jka} that appeared in the introduction. Notice that \eqref{skja8932}
also follows from \eqref{akiu32jkaAAA} by taking $\epsilon\to0$ and assuming that
${\cal S}$ of the roots $\zeta_j$ with $j\ne m$ vanish in this limit.

\section{Low energy spectrum in the scaling limit\label{sec3}}
\subsection{Preliminaries}
The low energy spectrum of the Hamiltonian of a 1D critical quantum spin chain at large system size
contains important information about the universal behaviour of the model
 \cite{Bloete:1986qm,Affleck1,Cardy:1986ie,Cardy84a}. As such,
upon the construction of the RG trajectories $\{|\Psi_L\rangle\}$, 
the corresponding eigenvalue of the Hamiltonian, $E(L)$, for $L\gg 1$  is one of the
first quantities that may be studied.
For the staggered six-vertex model with $U_q\big(\mathfrak{sl}(2)\big)$ invariant boundary conditions
in the critical regime $q=\re^{\ri\gamma}$ with $\gamma\in(0,\frac{\pi}{2})$
the energy 
of a certain class of low energy states was analyzed in the works 
\cite{Robertson:2020imc,Frahm:2021ohj}. The leading and sub-leading behaviour
is described as $E\asymp L e_\infty + f_\infty+o(1)$, where the
specific bulk energy $e_\infty$ and surface contribution to the energy $f_\infty$ 
were obtained in  ref.\cite{Frahm:2021ohj} within  the root density approach:
\bea\label{ajs1343}
e_\infty&=&-\int_{-\infty}^{+\infty}\rd \omega\,
\frac{\sinh(\frac{\pi \omega}{2n+4})}{\sinh(\frac{\pi\omega}{4})\,\cosh(\frac{n\pi\omega}{4n+8})}\\[0.2cm]
f_\infty&=&4\tan(\tfrac{\pi}{n+2})+\int^{\infty}_{-\infty}\rd \omega\, 
\frac{\cosh(\frac{\pi\omega}{4n+8})\,\sinh(\frac{\pi\omega (n-1)}{4n+8})}{\sinh(\frac{\pi\omega}{4})\,\cosh(\frac{n\pi\omega}{4n+8})}\,.\nonumber
\eea
Here and below the anisotropy $q$ is swapped for the parameter $n>0$ according to the relation
\be\label{askjb32nbas}
q=\re^{\ri\gamma}\,:\ \ \ \ \ \ \ \ \ \gamma=\frac{\pi}{n+2}\qquad\qquad\qquad (n>0)\, .
\ee
The next order term in the asymptotic expansion of the energy  goes as $1/L$.
Unlike $e_\infty$ and $f_\infty$, which are the same for all the low energy states, the 
coefficient for the $1/L$ 
correction  depends on the particular RG trajectory $\{|\Psi_L\rangle\}$ one is considering.
Among other things, it contains the conformal dimension characterising the scaling limit of that
state. In the works 
\cite{Robertson:2020imc,Frahm:2021ohj} the following asymptotic formula was proposed:
\be\label{asaaaaaakj90312}
E\asymp L\,e_\infty+f_\infty + \frac{\pi v_{\rm F}}{L}\bigg(\frac{p^2}{n+2}+\frac{b^2}{n}
-\frac{1}{12}+\texttt{d}\bigg)+o(L^{-1-\varepsilon})\ .
\ee
The Fermi velocity $v_{\rm F}$ is non-universal and depends on the overall multiplicative normalization
of the Hamiltonian. In our conventions for $\mathbb{H}$, one has
\be\label{ajs1344}
v_{\rm F}=2+\frac{4}{n}\ .
\ee
The term $p$ is shorthand for the expression
\be\label{asmn12bass}
p=\frac{1}{2}\,\big(2{\cal S}+1+\texttt{w}\,(n+2)\big)\,\qquad{\rm with}\qquad \texttt{w}=-1\ ,
\ee
where ${\cal S}$ is the $U_q\big(\mathfrak{sl}(2)\big)$ spin of $|\Psi_L\rangle$, while
the notation $\texttt{d}$ stands for `descendent' and is a non-negative integer
$\texttt{d}=0,1,2,\ldots\ $.  Also, the correction term $o(L^{-1-\varepsilon})$ contains an
infinitesimally small $\varepsilon>0$.
\medskip

The  quantity $b$
in the asymptotic formula \eqref{asaaaaaakj90312} requires special comment.
First of all, it turns out to depend on the system size $b=b(L)$.
In ref.\cite{Frahm:2021ohj} it was found that
$b(L)$ is related to the eigenvalue $B$ of the so-called `quasi-shift' operator $\mathbb{B}$
computed on $|\Psi_L\rangle$. This operator commutes with
the transfer-matrix and
coincides,  up to an overall factor,\footnote{The factor is given by $-\frac{q^{4L+2}}{1+q^{2}}\,f^2(\ri q^{-2})$
with $f$ being the function defined in \eqref{askjh21}.} with $\left(\mathbb{T}^{(0)}(\ri q^{-1})\right)^2$, see also ref.\,\cite{Frahm:2022gtk}.  The precise relation reads as
\be\label{ska823jhdsAAA}%
b(L)=\frac{n}{2\pi}\,\log\big(\sqrt{B}\,\big)\ .
\ee
Here $\sqrt{B}$ is given in terms of the Bethe roots corresponding to a Bethe state as
 \be\label{ska823jhds}
\sqrt{B}=\prod_{m=1}^{L-{\cal S}}\,\frac{\big(\zeta_m+\ri q^{-1}\big)\,\big(\zeta_m-\ri q\,\big)}{%
\big(\zeta_m-\ri q^{-1}\big)\,\big(\zeta_m+\ri q\,\big)}\ .
\ee
The asymptotic formula \eqref{asaaaaaakj90312} for  the low energy spectrum 
should be understood with $b$, therein,
 substituted for 
$b(L)$ obtained via \eqref{ska823jhdsAAA}.
\bigskip

\begin{figure}[t]
    \centering
    
    \begin{minipage}[b]{.49\linewidth}
    \scalebox{0.85}{
\begin{tikzpicture}
\node at (3,1) { $u$};
\draw (3,1.01) circle (0.25cm);
\node at (0,0) {\includegraphics[width=0.9\linewidth]{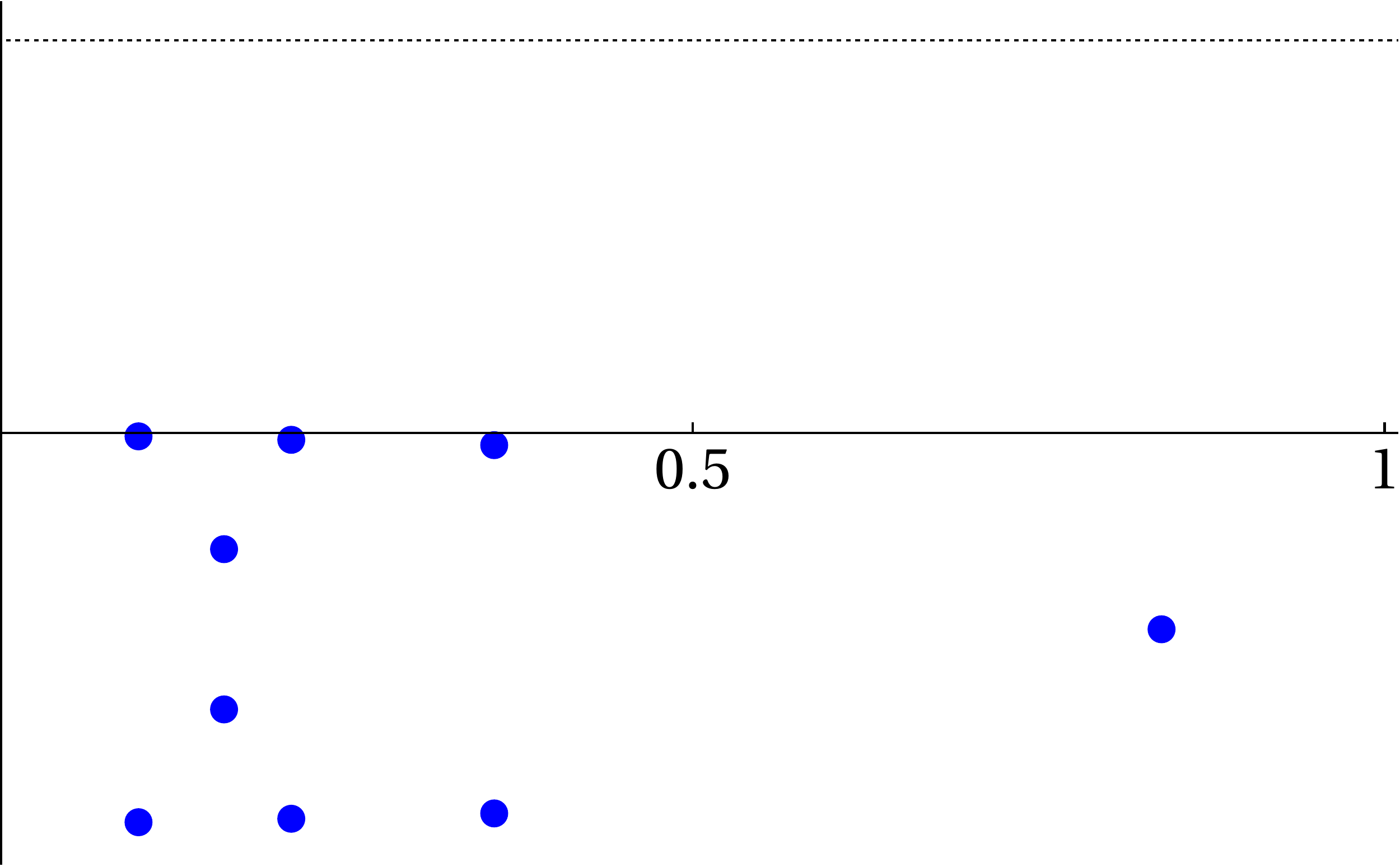}};
\node[anchor=west] at (1.75,2.35)   {$\Im m(u)=\frac{\pi}{2}$};
\end{tikzpicture}
}
 \end{minipage}
 \begin{minipage}[b]{.49\linewidth}
    \scalebox{0.85}{
\begin{tikzpicture}
\node at (0,0) {\includegraphics[width=0.9\linewidth]{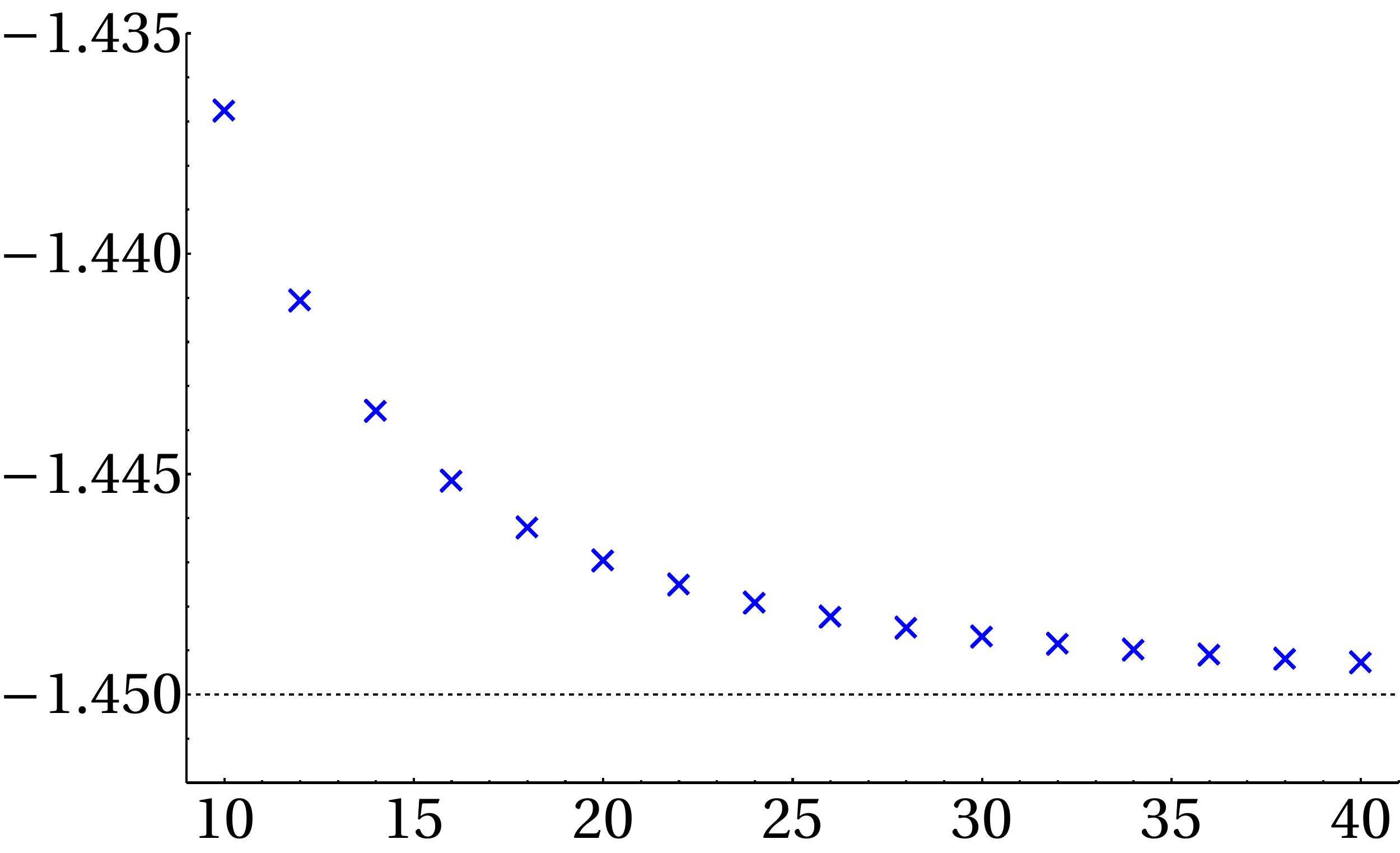}};
\node[anchor=west] at (3.6,-1.75)   {$L$};
\node[anchor=west] at (-3.5,2.5)   {$\Im m (b(L))$};
\end{tikzpicture}
}
 \end{minipage}
 
    \caption{\small Displayed is numerical data for an RG trajectory where $b(L)$ tends to $-\frac{\ri n}{2}$. The left panel depicts the pattern of Bethe roots in the complex $u=-\frac{1}{2}\log(\zeta)$ plane for the state $|\Psi_L\rangle$
with $L=10$. The right panel is a plot of $b(L)$ as a function of  $L$, where the dashed line represents its limiting value. Here $\mathcal{S}=1$ and  $n$, which parameterizes $q$ as in \eqref{askjb32nbas}, is given by $n=2.9$. \label{Dis_State_Minus_n_2}}
    % State number is 28
    % Saved under spectro/RG/Z28
\end{figure}
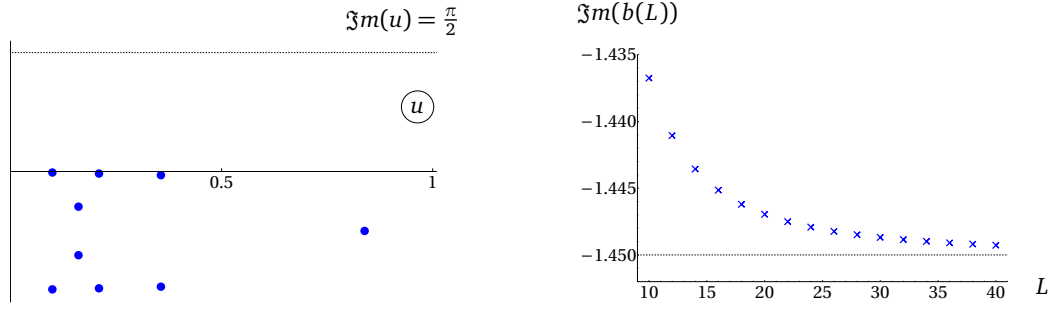

Some comment is required on the choice of the branch for  the logarithm in the expression for $b(L)$ 
\eqref{ska823jhdsAAA}. 
For all the trajectories we constructed, it turned out that  consistency with the
asymptotic formula for the energy \eqref{asaaaaaakj90312} requires that:
\be\label{ska823jhdsBBB}
-\frac{n}{2}<\Im m\big(b(L)\big) \le \frac{n}{2}\ .
\ee
The question of which of the boundaries 
 $\Im m\big(b(L)\big)=
 \pm\frac{n}{2}$  
to include in the domain of $b(L)$  does not matter for the following reason.
The only RG trajectories of the spin chain which were observed such that
$\Im m\big(b(L)\big)\to\pm \frac{\ri n}{2}$ as $L\to\infty$ had vanishing real part 
in the scaling limit. The typical
pattern of Bethe roots for one of these is depicted in Fig.\,\ref{Dis_State_Minus_n_2}. In this case, one notes that 
the asymptotic formula for the energy \eqref{asaaaaaakj90312} yields the same result
for $\lim_{L\to\infty} L\,(E-Le_\infty -f_\infty)$ regardless of whether  $\lim_{L\to\infty} b(L)$ coincides with
$-\frac{\ri n}{2}$ or 
$+\frac{\ri n}{2}$.
\medskip

The appearance of a  term like $b=b(L)$ in the $1/L$ corrections to $E(L)$,
that depends on the low energy state, 
was originally observed
in the context of 1D quantum spin chains with periodic boundary conditions imposed \cite{Essler:2005ag,Jacobsen:2005xz}. 
The problem of extracting the spectrum of conformal dimensions for these models
 was studied in detail in the later works \cite{Ikhlef:2011ay,Bazhanov:2019xvyA}.
The analysis therein carries over to the staggered six-vertex model
with open, $U_q\big(\mathfrak{sl}(2)\big)$
invariant boundary conditions we are considering.
As an illustration, let's discuss it first for
the class of states that were considered in the papers \cite{Frahm:2021ohj,Robertson:2020imc}.
\medskip

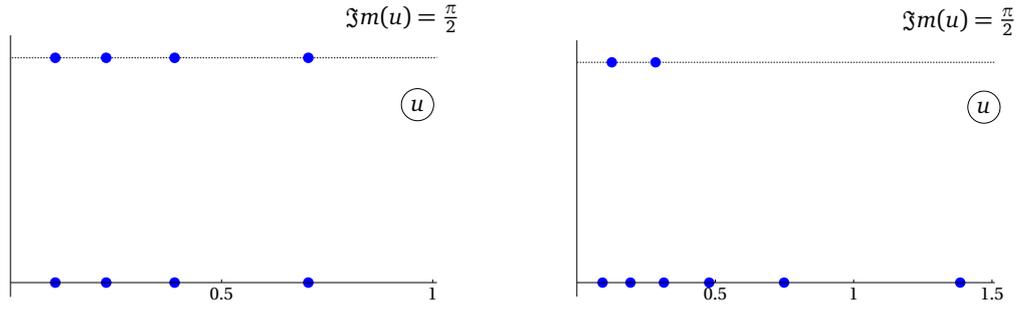
\begin{figure}[t]
    \centering
    
    \begin{minipage}[b]{.49\linewidth}
    \scalebox{0.85}{
\begin{tikzpicture}
\node at (3,1) { $u$};
\draw (3,1.01) circle (0.25cm);
\node at (0,0) {\includegraphics[width=0.9\linewidth]{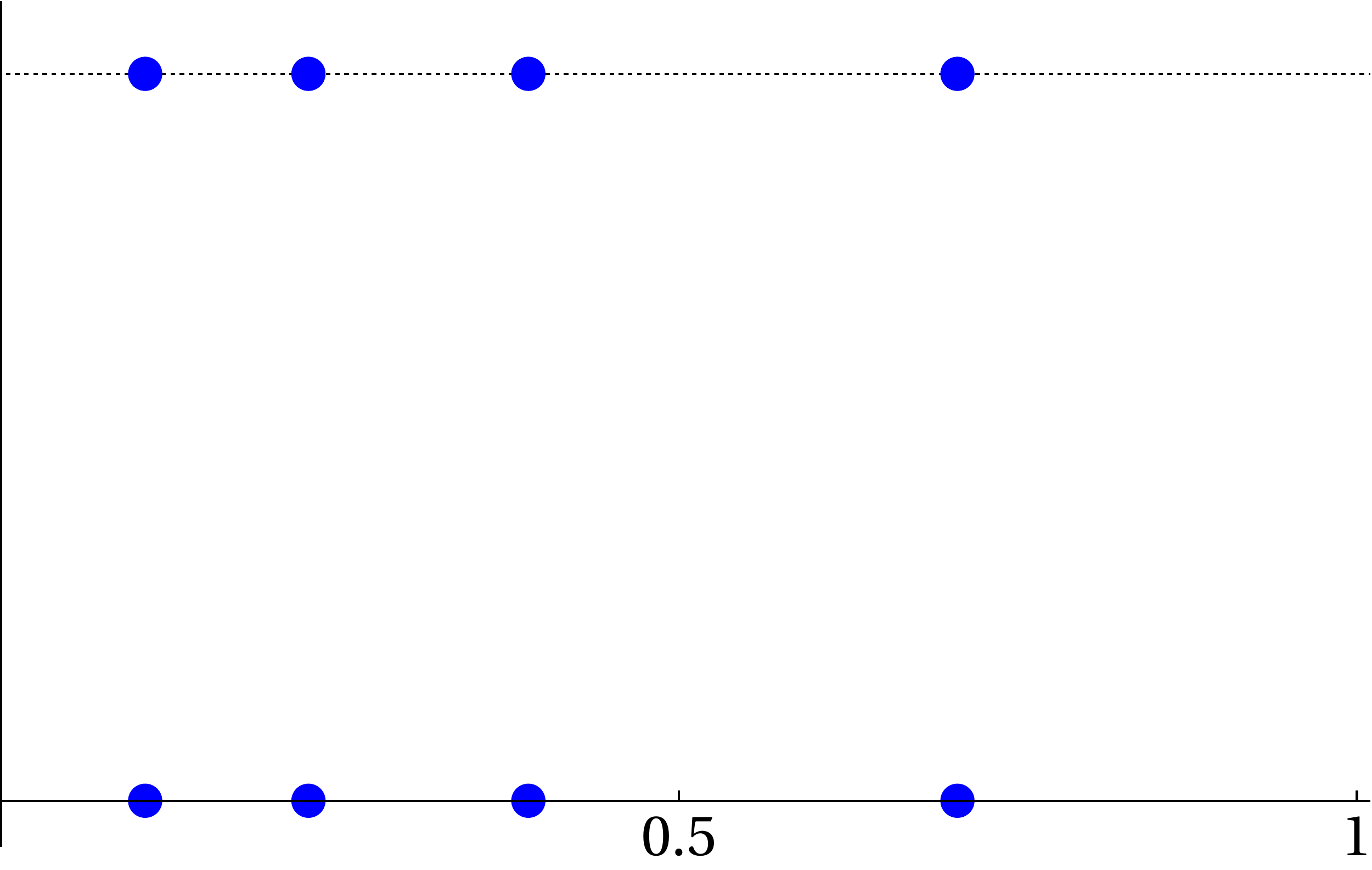}};
\node[anchor=west] at (1.75,2.35)   {$\Im m(u)=\frac{\pi}{2}$};
\end{tikzpicture}
}
 \end{minipage}
 \begin{minipage}[b]{.49\linewidth}
    \scalebox{0.85}{
\begin{tikzpicture}
\node at (3,1) { $u$};
\draw (3,1.01) circle (0.25cm);
\node at (0,0) {\includegraphics[width=0.9\linewidth]{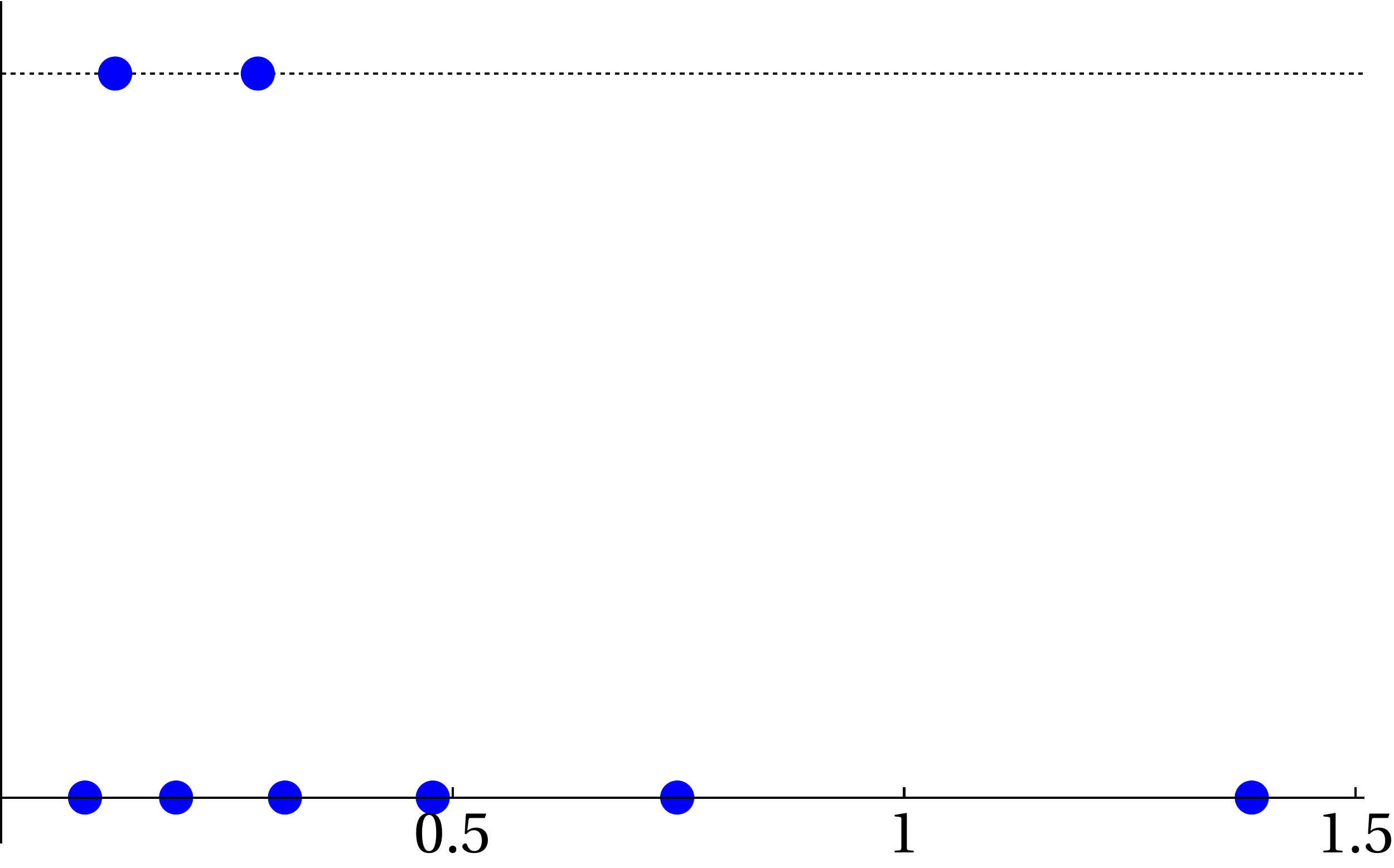}};
\node[anchor=west] at (1.6,2.35)   {$\Im m(u)=\frac{\pi}{2}$};
\end{tikzpicture}
}
 \end{minipage}
 
    \caption{\small On the left and right panels displayed are the Bethe roots in the complex  $u=-\frac{1}{2}\log (\zeta)$ plane for the ground state and an excited state, respectively, of the spin chain with $L=10$.  The excitation is built by disbalancing the number of roots on the two lines with respect to the ground state pattern. The states have total $U_q\big(\mathfrak{sl}(2)\big)$ spin $\mathcal{S}=2$ while $n=2.9$.}
    \label{BR_Classical_States}
\end{figure}

There exist solutions to the Bethe Ansatz equations \eqref{akiu32jka}
such that all the Bethe roots are real.
Different solution sets are distinguished by the integer $\mathfrak{m}$, which stands for the difference between
the number of roots lying on the positive and negative real axes (see Fig.\ref{BR_Classical_States}):
\be\label{asnbv12bas}
\{\zeta_m\}_{m=1}^{M}=\{\zeta_m^{(+)}\}_{m=1}^{M+\frac{\mathfrak{m}}{2}}\,\cup\, \{\zeta_m^{(-)}\}_{m=1}^{M-\frac{\mathfrak{m}}{2}}
\qquad\qquad {\rm with}\qquad\qquad  \zeta_m^{(\pm)}\gtrless 0
\qquad\qquad (M=L-{\cal S})\, .
\ee
Comparison of the energy computed from the solution set $\{\zeta_m\}$ via eq.\,\eqref{anms12i78} with that coming 
from the direct diagonalization of the Hamiltonian, for small lattice sizes, shows that the
corresponding Bethe  states are low
energy states of the spin chain provided $\mathfrak{m}\ll L$. The large $L$ asymptotic of the energy
turns out to obey \eqref{asaaaaaakj90312} with $\texttt{d}=0$ so long  as  there are no significant
gaps in the distribution of Bethe roots along the positive and negative rays $\zeta\lessgtr 0$. 
Computing the value of  $b(L)$ from the definition \eqref{ska823jhds}, \eqref{ska823jhdsAAA} one finds 
it to be a  real number, which turns out to satisfy the large $L$ asymptotic behaviour \cite{Frahm:2021ohj}
\be\label{masn891jhds}
b_{\mathfrak{m}}(L)=
\frac{\pi\mathfrak{m}}{2\log(L)}+O\big(1/(\log L)^2\big)\qquad\qquad\qquad (\mathfrak{m}\,-\,{\rm fixed})\, .
\ee
This way as 
$L$ becomes large, $\Delta b_{\mathfrak{m}}(L)=b_{\mathfrak{m}+2}(L)-b_{\mathfrak{m}}(L)\propto 1/\log L$, so that
the values of $b_{\mathfrak{m}}(L)$ are densely distributed in some segment of the real line. 
The latter is given by $(-b_{\mathfrak{m}_{\rm max}},+b_{\mathfrak{m}_{\rm max}})$, where
 $\mathfrak{m}_{\rm max}(L)\ll L$ is the maximum value of the integer $\mathfrak{m}$ such that
the state with Bethe roots \eqref{asnbv12bas} is still of low energy.
Assuming that  $\mathfrak{m}_{\rm max}$
grows faster than $\log(L)$ as $L\to\infty$, this segment becomes the entire real line in the scaling limit.
\medskip

When assigning an $L$ dependence $|\Psi_L\rangle$ 
to the class of states discussed above, it is tempting to keep the integer
$\mathfrak{m}$ fixed. Then, in view of formula \eqref{masn891jhds}, the
value of $b(L)$ would go to zero as $L\to\infty$. However, there is another way  of organizing the RG flow of the states. 
One may increase $\mathfrak{m}$ as $\sim\log(L)$ so that the  value of 
$b(L)$  tends to a finite, non zero limit as $L\to\infty$. Such an RG trajectory would be characterized by 
\be\label{aslj3298ds}
s=\slim_{L\to\infty} b(L)\,,
\ee
which can be arranged to be an arbitrary real number.  Here and below
we use the symbol $\slim$ for `scaling limit'
to emphasize that there is additional non-trivial input involved in taking the number of sites to infinity.
Let's compare the asymptotic formula for the energy \eqref{asaaaaaakj90312} for the RG trajectory with
the general  CFT prediction for a lattice system with open boundary conditions imposed \cite{Bloete:1986qm,Cardy84a}:
\be
E\asymp L\,e_\infty+f_\infty + \frac{\pi v_{\rm F}}{L}\bigg(\Delta-\frac{c}{24}\bigg)+o(L^{-1-\varepsilon})\ ,
\ee
where $\Delta$ is the conformal dimension, while $c$ stands for the central charge.
One finds that
\be\label{sa8721hdsb}
\Delta-\frac{c}{24}=\frac{p^2}{n+2}+\frac{s^2}{n}
-\frac{1}{12}+\texttt{d}
\ee
with $\texttt{d}=0$.
Thus we conclude that the spectrum of scaling dimensions develops a continuous component
labeled by the parameter $s\in\mathbb{R}$.
\medskip

It should be mentioned that in the work \cite{Frahm:2021ohj} RG trajectories were also constructed for
which $b(L)$ tends to a pure imaginary number in the scaling limit. Among them is the low energy state
mentioned in the paragraph containining formula \eqref{ska823jhdsBBB}, whose typical pattern of
Bethe roots in the complex $u=-\frac{1}{2}\log(\zeta)$ plane is depicted in Fig.\,\ref{Dis_State_Minus_n_2}.
Moreover, the authors of ref.\cite{Frahm:2021ohj} propose that the imaginary values of $s$ that appear
in the scaling limit of the lattice model must satisfy the condition
\be\label{askhbaaa23}
s=\pm\ri \,\Big(-p-\frac{1}{2}-a\,\Big)\qquad\qquad{\rm with}\qquad\qquad a=0,1,2,\ldots <-p-\tfrac{1}{2}\ 
\qquad\qquad (\texttt{d}=0)\, .
\ee

\subsection{ODE/IQFT correspondence\label{sec32}}
In order to characterize the scaling limit of a lattice system,
where the spectrum of scaling dimensions possesses a continuous component, mathematical techniques are required
that go beyond those used in the works \cite{Robertson:2020imc,Frahm:2021ohj}.
One of these is
an approach to the study of the scaling limit that is based on the so-called ODE/IQFT correspondence 
\cite{Voros:1994,Dorey:1998pt,Bazhanov:1998wj,Bazhanov:2003ni}. We found that the ODEs which 
describe the scaling limit of the staggered six-vertex model with $U_q\big(\mathfrak{sl}(2)\big)$
invariant BCs lie within the class of differential equations 
considered in refs.\cite{Bazhanov:2019xvy,Bazhanov:2019xvyA} in the context of the lattice system
with quasi-periodic BCs. This  made the analysis performed in our work possible. In addition, it allowed us to transfer over many previous results concerning the differential equations.
Here we briefly discuss the application of the  ODE/IQFT correspondence to the study of the scaling limit of the spin chain, while
referring the reader to ref.\cite{Bazhanov:2019xvyA} for technical details.
\medskip

The primary Bethe states, i.e.,  the RG trajectories 
where the energy obeys the asymptotic formula \eqref{asaaaaaakj90312} with $\texttt{d}=0$, 
are labeled by the $U_q\big(\mathfrak{sl}(2)\big)$ spin ${\cal S}$ and the RG invariant $s$ 
defined in eq.\,\eqref{aslj3298ds}. The ODE/IQFT correspondence implies a relation
between the scaling limit of the eigenvalue of the $Q$ operator for $|\Psi_L\rangle$ 
and the spectral determinant of the ODE:
\bea\label{aisausau}
\Bigg[\,-\frac{\rd ^2}{{\rd z}^2}+\frac{p^2-\frac{1}{4}}{z^2}+\frac{2\ri  s}{z}+1+\mu^{-2-n}\ z^n\,
\Bigg]\, \psi=0\ .
\eea
Here $p$ is given in terms of ${\cal S}$ as in eq.\,\eqref{asmn12bass}, while $n>0$ is related to the anisotropy 
parameter of the spin chain, see \eqref{askjb32nbas}. In addition, in taking the scaling limit
the parameter $\zeta$ entering into the $Q$ operator is assigned a certain $L$ dependence such that
$\zeta\propto  L^{-\frac{n}{n+2}}$.  Then $\mu$ appearing in the ODE \eqref{aisausau}  is given by 
\be
\mu=-\ri\,\big(L/L_0)^{\frac{n}{n+2}}\,\zeta
\ee
with
\be\label{kjsa8sss9321}
L_0=\frac{\sqrt{\pi}\,\Gamma\big(1+\frac{1}{n}\big)}{4\Gamma\big(\tfrac{3}{2}+\frac{1}{n}\big)}\, .
\ee
The spectral determinant $D(\mu\,|\,p,s)$ is defined in the following way. 
One specifies a solution to the differential equation by its behaviour in the
vicinity of the singular point $z=0$:
\be\label{psidef1}
\psi_{p}(z)\to 
z^{\frac{1}{2}+ p} \ \ \ \ \ \ \  {\rm as} \ \ \ \ \ \ \ z\to 0  \ .
\ee
For large $z$ the term $\mu^{-2-n}\,z^n$ in \eqref{aisausau} 
becomes dominant and one can define another solution through the $z\to+\infty$  asymptotic:
\be\label{Ajsh13b3}
\Xi(z)\asymp  \Big(\frac{z}{\mu}\Big)^{-\frac{n}{4}}\,\exp\bigg[-\frac{2}{n+2}\  \Big(\frac{z}{\mu}\Big)^{\frac{n}{2}+1}\, 
{}_2F_1\Big( -\tfrac{1}{2}, -\tfrac{n+2}{2n},  
\tfrac{n-2}{2n}\,\big|-\mu^{n+2}\, z^{-n}  \Big)\ +\ o(1)\,\bigg]
\ee
with  ${}_2F_1(a,b,c|z)$ being the usual Gauss hypergeometric function and   we make the technical assumption that $\mu>0$ and $n\ne\frac{2}{2k-1}$
with $k=1,2,\ldots\ $. 
The spectral determinant $D(\mu)=D(\mu\,|\,p,s)$ is given by
\be\label{Ddef1a}
D(\mu)=\sqrt{\pi}\ 
(n+2)^{-\frac{2p}{n+2}-\frac{1}{2}}\ \mu^{- p+\frac{1}{2}}\ 
\frac{W[\,\psi_{p},\Xi\,]
}{\Gamma(1+\frac{2p}{n+2})}\,,
\ee
where $W[\,\psi_{ p},\Xi\,]=\Xi\,\partial_z\psi_{ p}-\psi_{ p}\,\partial_z\Xi$ is the Wronskian.
The overall factor  has been chosen to ensure the normalization
\be\label{iaosid91209}
D(0)= 1\ .
\ee
It should be mentioned that  the procedure for computing the spectral determinant
based on formula \eqref{Ddef1a} 
with the solutions $\psi_p$ and $\Xi$ obtained via a numerical
integration of the ODE \eqref{aisausau} 
 works literally only for $\Re e(p)\ge 0$. Nevertheless $D(\mu\,|\,p,s)$ 
turns out to be a meromorphic function of $p$ and can be defined  
for generic complex values of this variable via analytic continuation.
\medskip

Rather than considering the eigenvalue of $\mathbb{Q}^{(0)}(\zeta)$ \eqref{Eigenvalue_Q_0_BR}  for a primary Bethe state,
we instead discuss the scaling limit of 
\be
A(\zeta)=\prod_{j=1}^{L-{\cal S}}\,(1-\zeta/\zeta_j)\,(1-\zeta\zeta_j)\,.
\ee
It does not involve the  overall factor $\zeta^{\cal S}$  and
the normalization has been  imposed such that $A(0)=1$.
Then, the scaling relation between $A(\zeta)$ and the spectral determinant reads as
\bea\label{as56d1a}
\slim_{L\to \infty\atop b(L)\to s}
G^{(L)}\Big(-\mu^2\Big|\tfrac{2}{n+2}\Big)\ A\Big(\big(L/L_0\big)^{-\frac{n}{n+2}}\ \ri\mu \Big)={D}(\mu)\, .\qquad
\eea
Here the function $G$ has been chosen to ensure the convergence of the limit and is given by
    \begin{align}
        G^{(L)}(E\,|g)=\exp\left( \sum^{\big[\frac{1}{2(1-g)}\big] }_{m=1} \frac{(-1)^m\,L}{m\cos(\pi m g )} \left(\frac{L}{L_0}\right)^{2m(g-1)}E^m\right)\,,
    \end{align}
(this should be compared with formula (5.48) in the work \cite{Bazhanov:2019xvyA}), where the brackets $[\ldots]$ stand
for the integer part, while $L_0$ is the same as in \eqref{kjsa8sss9321}.
As above, the technical assumption
\be
 n\neq \frac{2}{2k-1}
\ee
with $k=1,2,\ldots$ is being made,
see ref.\cite{Bazhanov:2019xvyA} for details concerning the case 
 $n=\frac{2}{2k-1}$.
\medskip

For the RG trajectories, where $\texttt{d}$ entering into the 
asymptotic formula for the energy \eqref{asaaaaaakj90312}
is greater than zero, the scaling relation \eqref{as56d1a}
is modified as follows. The l.h.s. remains the same, while for the r.h.s. one takes 
$D(\mu)$ to be the spectral determinant
for the differential equation:
\bea\label{aisausauAAA}
\Bigg[\,-\frac{\rd ^2}{{\rd z}^2}+\frac{p^2-\frac{1}{4}}{z^2}+\frac{2\ri  s}{z}+1+
\sum_{a=1}^\texttt{d}\bigg(\frac{2}{(z-w_a)^2}+\frac{n}{z(z-w_a)}\bigg)+\mu^{-2-n}\ z^n\,
\Bigg]\, \psi=0\ .
\eea
Here 
 \be
 \pmb{w}=(w_1,\ldots,w_\texttt{d})
 \ee
are not arbitrary parameters. They are restricted by the condition that
any solution $\psi(z)$ of the differential equation must be single valued in the vicinity of $z=w_a$.
This leads to the coupled algebraic system:
\bea\label{sksksk1}
4 n\, w_a^2\!\!&+&\!\!8\ri s\, (n+1)\, w_a-(n+2)\ \big((n+1)^2-4p^2\,\big)\\[0.2cm]
&+&\!\!
4\ \sum_{b\not=a}^{\texttt{d}}\frac{w_a\, \big(\,
 (n+2)^2\, w_a^2- n(2n+5)\, w_a w_b + n(n+1)\, w_b^2\,\big)}{(w_a-w_b)^3}=0\  \ \ \ \ \ \ \ \ (a=1,\ldots,
\texttt{d})\,.\nonumber
\eea
For generic $n$, $s$ and $p$ the number of solutions $\pmb{w}=\{w_a\}_{a=1}^\texttt{d}$, up to permutations of the $w_a$'s,
is given by ${\rm par}_2(\texttt{d})$ -- the number of bipartitions of $\texttt{d}$. The generating function for
this combinatorial quantity reads as:
\be\label{mas873hsd}
\sum_{\texttt{d}=0}^\infty{\rm par}_2(\texttt{d})\, \texttt{q}^\texttt{d}=
\prod_{j=1}^\infty \frac{1}{(1-\texttt{q}^j)^{2}}\ .
\ee
For applications to the staggered six-vertex model with $U_q\big(\mathfrak{sl}(2)\big)$
invariant BCs $p$ is not generic, but should be taken as in \eqref{asmn12bass}, i.e.,
$2p=2{\cal S}+1-(n+2)$. Then, it turns out that the number of solutions of the coupled equations \eqref{sksksk1}
is typically less than ${\rm par}_2(\texttt{d})$. To explain this phenomenon, 
let's replace $p$ with
$p_{\varepsilon}=p+\varepsilon^{2{\cal S}+1}$ where $0<\varepsilon\ll 1$.
Of the ${\rm par}_2(\texttt{d})$ solution sets of \eqref{sksksk1}  with $p\mapsto p_\varepsilon$ 
there exist those where
\be
w_a=O(\varepsilon)
\,\qquad\qquad {\rm for}\qquad\qquad a=1,2,\ldots,2{\cal S}+1\ .
\ee
The other variables $\{w_a\}_{a=2{\cal S}+2}^\texttt{d}$ tend to a finite,
nonvanishing limit as $\varepsilon\to 0$. Their limiting values obey
\eqref{sksksk1} with the replacements $p\mapsto {\cal S}+\frac{1}{2}+\frac{1}{2}\,(n+2)$
and $\texttt{d}\mapsto \texttt{d}-2{\cal S}-1$. In counting the solution
sets of the algebraic system on $w_a$ with $2p=2{\cal S}+1-(n+2)$ we 
only consider those to be admissible where none of the $w_a$ are zero.
It is easy to see that
\be\label{ksajnb389jhdsa}
N(\texttt{d}\, |\, \mathcal{S}):=\#\ {\rm of\ solution\ sets\ of\ \eqref{sksksk1}\ with}
\ p\ {\rm as\ in\ \eqref{asmn12bass}}= {\rm par}_2(\texttt{d})-{\rm par}_2(\texttt{d}-2{\cal S}-1)
\ee
 (we take by definition ${\rm par}_2(\texttt{d})=0$ when  its argument is a negative integer).
\medskip

We suppose that for a given trajectory $\{|\Psi_L\rangle\}$ with RG invariants
${\cal S}$, $s$ and $\texttt{d}$ there exists a solution set $\pmb{w}$
of \eqref{sksksk1} 
such that the scaling relation \eqref{as56d1a} holds true with
$D(\mu)=D(\mu\,|\,\pmb{w},p,s)$ being the spectral determinant for the differential
equation \eqref{aisausauAAA}. Note that eqs.\,\eqref{psidef1}-\eqref{Ddef1a}
for the definition of $D(\mu)$ still remain valid
 since the inclusion of the extra sum  in the ODE  has no impact on
 the leading asymptotics of $\psi_p$ and $\Xi$. 
\medskip

Unfortunately, we do not
know of a way of rigorously proving the above statement. We checked it numerically
for a variety of cases using the so-called sum rules. The analysis is not included here,
as it is essentially the same as that presented in section  11 of the work \cite{Bazhanov:2019xvyA}
concerning the staggered six-vertex model with quasi-periodic BCs.
We wish, however, to mention a scaling relation involving the products over the Bethe roots,
\be\label{skahg238}
\Pi_\pm=\prod_{m=1}^{L-{\cal S}}q\,
\big(\zeta_m\pm\ri q^{-1}\big)\,
\big(\zeta_m^{-1}\pm\ri q^{-1}\big)\,,
\ee
that will become important later.
It involves
the coefficients $\mathfrak{C}^{(\pm)}_{p,s}=\mathfrak{C}^{(\pm)}_{p,s}(\pmb{w})$, which occur in the large $\mu$
asymptotic expansion of $D(\mu)$:
\be\label{Apeq1}
D(\mu\,|\,\pmb{w},p,s)\,\asymp\,  { \mathfrak C}_{p,s}^{(\pm)}(\pmb{w})\ 
 \big(\pm \mu\big)^{\pm\frac{\ri (n+2) s}{n}-p}\,\exp\bigg(\, 
\frac{2L_0}{\cos(\frac{\pi}{n})}\ 
\big(\pm \mu\big)^{\frac{n+2}{n}}+o(1)\,\bigg) \ \ \ \, {\rm for}\ \ \ \  \Re e(\pm \mu)>0
 \ee
(again, we assume that $n\ne\frac{2}{2k-1}$ with $k=1,2,\ldots\ $). 
For the case when $\texttt{d} =0$, the coefficients are given by
\be\label{nsanb3vb23AAA}
\mathfrak{C}_{p,s}^{(0,\pm)}= 
\sqrt{\frac{2\pi}{n+2}}\ \ \ 2^{-p\pm\frac{\ri(n+2)s}{n}}\ 
(n+2)^{-\frac{2p}{n+2}}\ 
\frac{\Gamma(1+2p)}{\Gamma(1+\frac{2p}{n+2})\,\Gamma(\frac{1}{2}+p\pm\ri s)}\ \,.
\ee
In general,
\be\label{nsanb3vb23BBB}
\mathfrak{C}^{(\pm)}_{p,s}
(\pmb{w})=\mathfrak{C}_{p,s}^{(0,\pm)}\,\check{\mathfrak{C}}^{(\pm)}_{p,s}(\pmb{w})\,,
\ee
where $\check{\mathfrak{C}}^{(\pm)}$ are normalized to be one for $\texttt{d}=0$.
A closed form expression for 
 $\check{\mathfrak{C}}^{(\pm)}$ for general $\texttt{d}=0,1,2,\ldots$ was obtained in ref.\cite{Kotousov:2019nvt}
and, for the reader's convenience, is reproduced in this paper in Appendix \ref{AppB}.
The scaling relation reads as
\be\label{mn89sdjhsa}
\Pi_\pm\asymp\frac{C}{2\cos(\frac{\pi}{n+2})}\,\re^{\pm\frac{\pi}{n}s}
\,{ \mathfrak C}_{p,s}^{(\pm)}(\pmb{w})
 \,
 \left(\frac{L}{L_0}\right)^{-\frac{np}{n+2}\pm\ri s}\left(\frac{4n}{n+2}\right)^{L}\,\big(1+O(L^{-\epsilon})\big)\ .
\ee
It involves a non-universal constant $C$, which is expressed in terms of 
\be
    \hat{\tau}(\omega)=-\frac{1}{4\pi}\ \frac{\sinh(\frac{\pi(n-1)}{4(n+2)}\,\omega)}{\sinh(\frac{\pi \omega}{4(n+2)}) \cosh (\frac{n\pi \omega}{4(n+2)})}
\ee
and the Lerch transcendent
\be
\Phi(z,s,a)=\sum_{m=0}^\infty\frac{z^m}{(m+a)^s}\, .
\ee
Namely,
\be
C=\exp\Bigg(2\int_{-\infty}^\infty\rd \omega\,\bigg(
\frac{\hat{\tau}(\omega)}{\omega}\,\Big(\Im m\Big[\re^{\frac{2\ri\pi}{n+2}}\,
\Phi\big(\re^{-\frac{\ri n \pi}{n+2}}\,,\,1\,,\,1-\tfrac{\ri\omega}{4}\big)\Big]-\frac{\pi}{n+2}-\frac{2}{\omega}\,\Big)
-\frac{n-1}{2\pi\omega^2}\bigg)\Bigg)\ .
\ee
Also, the notation  $O(L^{-\epsilon})$ with some $\epsilon>0$ means that the correction terms
fall off faster than any power of the logarithm of $L$.
The asymptotic formula \eqref{mn89sdjhsa} is the analogue  for the lattice system with open $U_q\big(\mathfrak{sl}(2)\big)$
invariant BCs
of a  product rule presented  in 
the work \cite{Bazhanov:2019xvyA}, see (11.19) therein.
%Some of the data that was used for its
%numerical verification is displayed in Fig.\,\ref{}.

\subsection{Quantization condition}
Consider the problem of computing the spectrum of conformal dimensions
of the critical lattice system and the classification of the   space of
states ${\cal H}$ appearing in the scaling limit. 
The asymptotic relation for the energy \eqref{asaaaaaakj90312}, together
with the numerical analysis from refs.\cite{Robertson:2020imc,Frahm:2021ohj} 
concerning the large $L$ behaviour of $b(L)$ for a class of states, shows that
${\cal H}$ contains
 two components. 
There is a  `continuous component', where  $s=\slim_{L\to\infty}b(L)$ 
parameterising the conformal dimensions
as in  eq.\,\eqref{sa8721hdsb} may take any real value, as well as a `discrete' one
for which $s$ belongs to a finite set of pure imaginary numbers. In the next section
a full description of these linear spaces will be given. Among other things, 
this includes the admissible values of pure imaginary $s$ as well as the density of
states characterizing the continuous spectrum.
The results are based on an analysis of the so-called `quantization condition' for $b(L)$,
which we shall obtain below.
\medskip

The key observation is that the square root of the eigenvalue of the quasi-shift operator
\eqref{ska823jhds},
used in the computation of $b(L)$ \eqref{ska823jhdsAAA}, may be expressed as
\be
\sqrt{B}=(-1)^{L-{\cal S}}\ \frac{\Pi_+}{\Pi_-}\ .
\ee
Here $\Pi_\pm$ stand for the products over the Bethe roots defined in  \eqref{skahg238}.
Let us substitute these products for their asymptotics \eqref{mn89sdjhsa} with $s$
replaced by the `running coupling' $b(L)$. Upon rearranging and making use of
eq.\,\eqref{ska823jhdsAAA}, one finds
\begin{subequations}\label{kajs8712jhds1}
\be
\bigg(\frac{L}{L_0}\bigg)^{2\ri s}\re^{\frac{\ri}{2}
\delta(\pmb{w},p,s)}\Big|_{s=b(L)}=\sigma+O(L^{-\epsilon})
\ee
with
\be\label{asnbbvb32vbAAAAv}
 \re^{\frac{\ri}{2}
 \delta(\pmb{w},p,s)}=\frac{\mathfrak{C}^{(+)}_{p,s}(\pmb{w})}{\mathfrak{C}^{(-)}_{p,s}
 (\pmb{w})}
\ee
and 
\be\label{m23n8723u21}
\sigma=(-1)^{L-{\cal S}}\ .
\ee
\end{subequations}
The relation \eqref{kajs8712jhds1}, 
which will be henceforth referred to as the quantization condition, is interpreted in the following way.
Given an RG trajectory
$\{|\Psi_L\rangle\}$ one computes $p$ from the value of the $U_q\big(\mathfrak{sl}(2)\big)$ spin
${\cal S}$ via the definition \eqref{asmn12bass} as well as the sign factor $\sigma=(-1)^{L-{\cal S}}$. 
The latter is kept fixed along $\{|\Psi_L\rangle\}$ since, in the construction of RG trajectories,
 $L$ is always increased by two (see section \ref{sec2}). Then $b(L)$ computed from the Bethe roots according to
 eqs.\,\eqref{ska823jhds} and \eqref{ska823jhdsAAA} obeys \eqref{kajs8712jhds1}
for some solution set $\pmb{w}=\{w_a\}_{a=1}^\texttt{d}$ of the algebraic system \eqref{sksksk1} with $s$ replaced by $b(L)$.
The `phase shift' $\delta$ is given in terms of the coefficients
$\mathfrak{C}^{(\pm)}_{p,s}(
\pmb{w})$ \eqref{nsanb3vb23BBB}, which were introduced in the previous subsection.
For the primary Bethe states with $\texttt{d}=0$, one has
\be\label{askjjh1287sd}
 \re^{\frac{\ri}{2}\delta(\emptyset,p,s)}=
2^{\frac{2\ri(n+2)s}{n}}\ \frac{\Gamma(\frac{1}{2}+p-\ri s)}{\Gamma(\frac{1}{2}+p+\ri s)}\qquad
\qquad\qquad (\texttt{d}=0)\, .
\ee
For $\texttt{d}=1,2,3,\ldots$ one must make use of eqs.\,\eqref{nsanb3vb23AAA},\,\eqref{nsanb3vb23BBB} together
with the explicit formula for $\check{\mathfrak{C}}^{(\pm)}_{p,s}(\pmb{w})$ as a function of $p$, $s$ and
$\pmb{w}=\{w_a\}_{a=1}^\texttt{d}$ contained in Appendix \ref{AppB}.
\medskip

\begin{figure}
\centering
\begin{tikzpicture}
\node at (0,0) {\includegraphics[width=8cm]{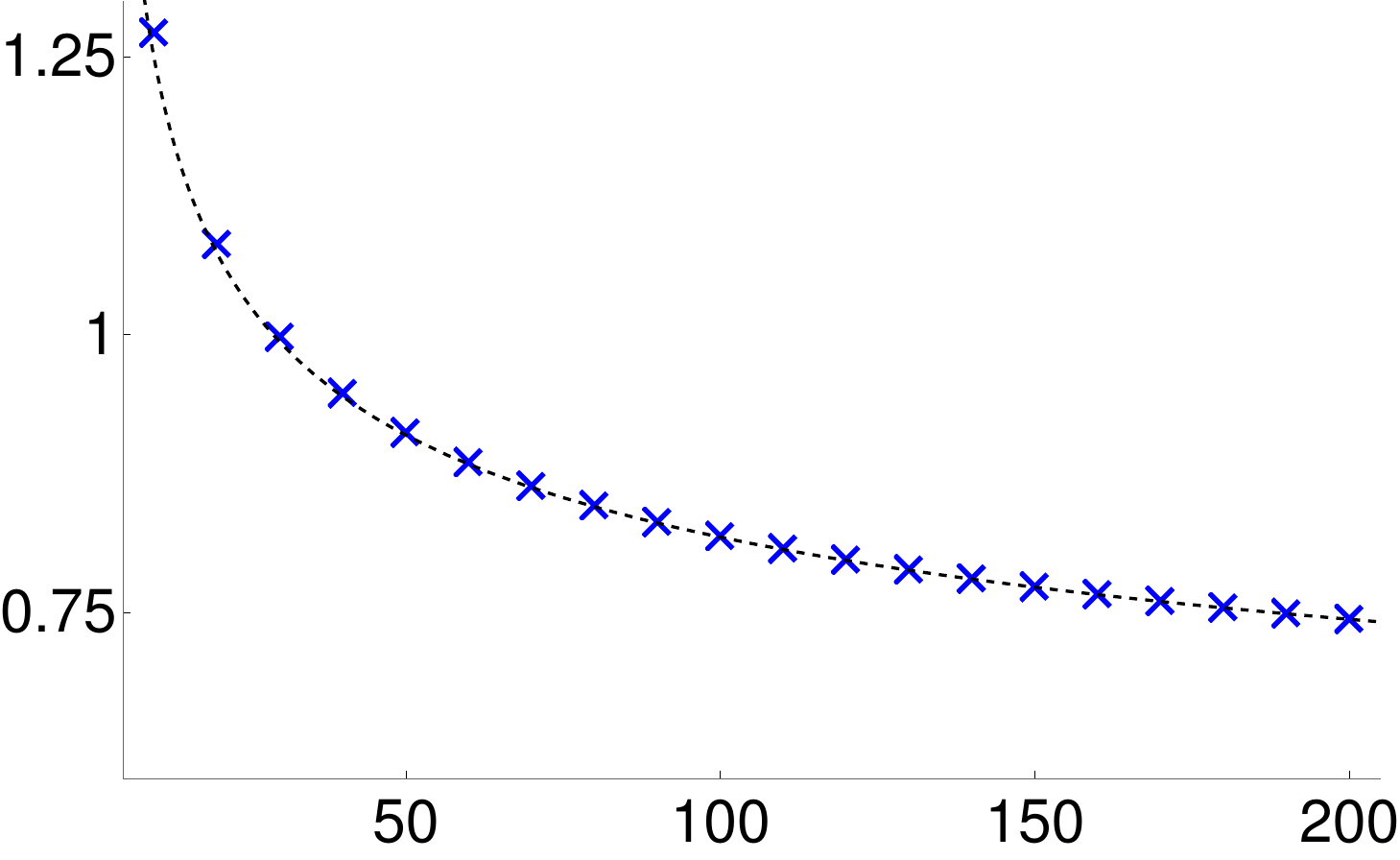}};
\node at (-3.3,2.85) {\large $b(L)$};
\node at (4.4,-2.0) {\large $L$};
\end{tikzpicture}
\caption{\small%
The numerical data comes from an RG trajectory $\{|\Psi_L\rangle\}$, where the representative
state for $L=10$ is the one whose pattern of Bethe roots is displayed on the right panel of fig.\,\ref{BR_Classical_States}.
In particular, it has $\texttt{d}=0$; total $U_q\big(\mathfrak{sl}(2)\big)$ spin ${\cal S}=2$; while the integer $\mathfrak{m}$ ---
the difference between the number of roots lying on the real line and the line $\Im m(u)=\frac{\pi}{2}$ in the complex
$u$ plane --- is held fixed along the flow to be $\mathfrak{m}=4$. The crosses depict the numerical values of 
$b(L)=b_{\mathfrak{m}}(L)$
for different $L$ which were computed from the Bethe roots corresponding to $|\Psi_L\rangle$ 
via formulae \eqref{ska823jhdsAAA} and \eqref{ska823jhds}. The  dashed line gives the predictions coming from
the quantization condition \eqref{kaaaasj9812}
with $\mathfrak{m}=4$ and  the correction terms ignored. Note that the branch of the logarithm was fixed
by requiring that the l.h.s. of  \eqref{kaaaasj9812} is a continuous function for real $b_{\mathfrak{m}}$ which
vanishes at $b_{\mathfrak{m}}=0$.
\label{sdsdjhsjdhsjdaaa}}
\end{figure}

Let's take a moment to discuss  the quantization condition \eqref{kajs8712jhds1} for the primary Bethe states
in the context of the results of the previous works \cite{Frahm:2021ohj,Robertson:2020imc}.
We start with the asymptotic  \eqref{masn891jhds} for $b(L)$ that was observed for
a class of RG trajectories labelled by the integer $\mathfrak{m}$.  In this case, it is
useful to take the logarithm of both sides of formula \eqref{kajs8712jhds1} with the 
phase shift as in \eqref{askjjh1287sd}
and write it in the form:
\be\label{kaaaasj9812}
2 b_\mathfrak{m}\log\bigg(\frac{L}{L_0}\bigg)-\ri
\log\bigg[2^{\frac{2\ri(n+2)b_{\mathfrak{m}}}{n}}\
 \frac{\Gamma(\frac{1}{2}+p-\ri b_\mathfrak{m})}{\Gamma(\frac{1}{2}+p+\ri b_\mathfrak{m})}\bigg]=\pi\mathfrak{m}+O(L^{-\epsilon})\ .
\ee
For the class of states we are considering
 $b_{\mathfrak{m}}(L)$   goes to zero as $L\to\infty$. As a result, the second term 
in the l.h.s. of the above relation containing the $\Gamma$-functions 
also tends to zero and, to a first approximation, can be ignored.
This way one obtains \eqref{masn891jhds}. Formula \eqref{kaaaasj9812}  provides a refinement to the large $L$
asymptotic behaviour of $b_{\mathfrak{m}}(L)$
which takes into account all power law corrections in $1/\log(L)$. To demonstrate its  accuracy,
some numerical data obtained from the Bethe roots for a primary Bethe state $|\Psi_L\rangle$ is compared with the
predictions coming from the quantization condition in Fig.\,\ref{sdsdjhsjdhsjdaaa}.
\medskip

Another possibility of how  \eqref{kajs8712jhds1} could be satisfied for $L\gg 1$  is if
$b(L)$ approaches a singularity of the phase shift. The explicit formula \eqref{askjjh1287sd},
valid for $\texttt{d}=0$,
shows that these  occur for pure imaginary $s$  when
$\frac{1}{2}+p\pm\ri s$ is a positive integer. If the imaginary part of $b(L)$ is positive,
then the vanishing of the first term in the l.h.s. of \eqref{kajs8712jhds1} may be compensated if
$b(L)$ tends to a pole of $\re^{\frac{\ri}{2}\delta}$, i.e.,
\be
\slim_{L\to\infty}b(L)=s\qquad \qquad {\rm with}\qquad \qquad s=\ri\big(-p-\tfrac{1}{2}-\ell\big)
\qquad\qquad \big(\Im m(b(L))>0\big)
\ee
and $\ell=0,1,2,\ldots\ $. This is the same as eq.\,\eqref{askhbaaa23} with the sign factor chosen to be `$+$'.
The
upper bound on $\ell$ in that equation ensures  the condition $\Im m(b(L))>0$.
The minus version of the relation is deduced from \eqref{kajs8712jhds1} 
by means of similar arguments.
\medskip

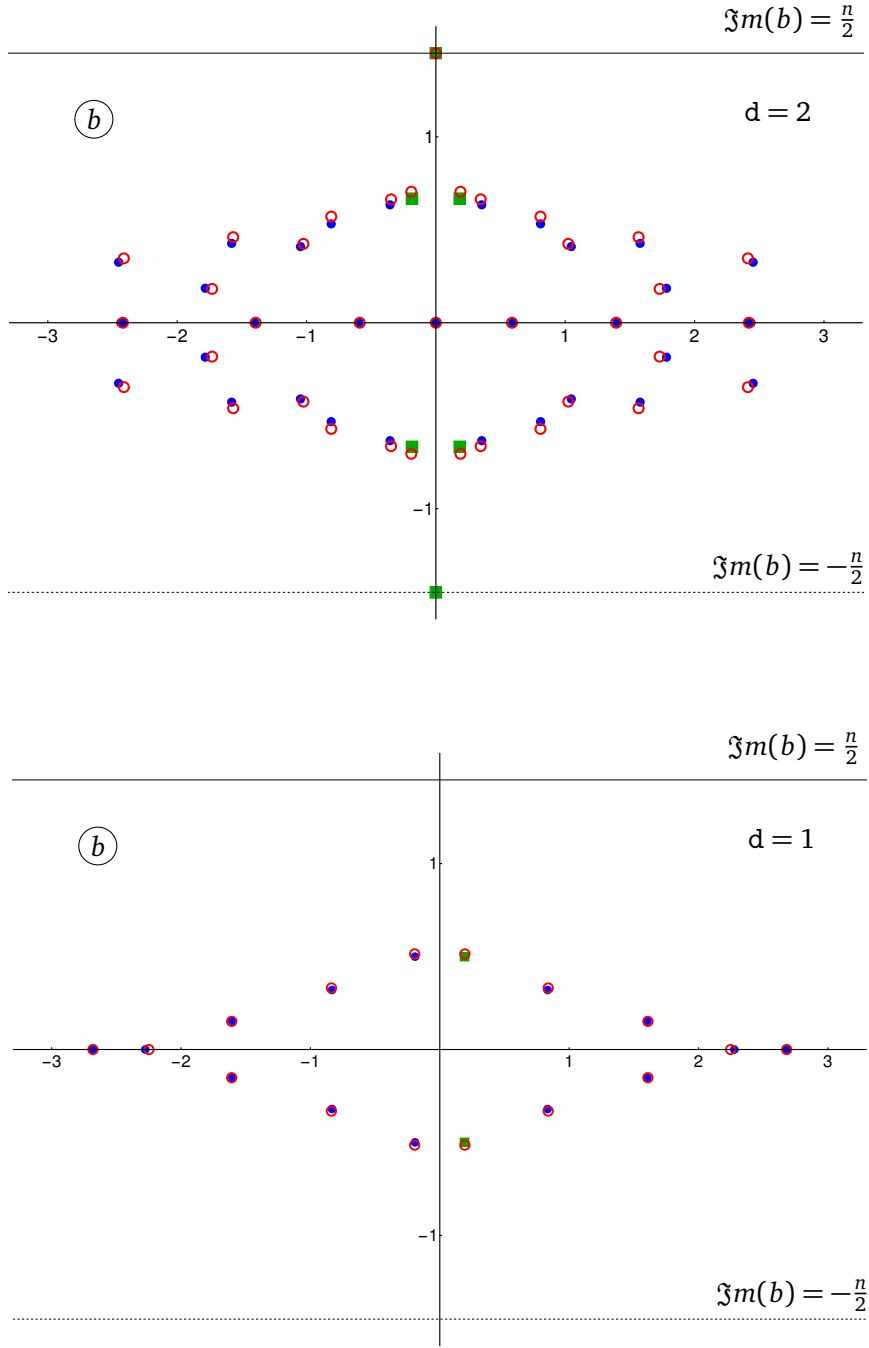
\begin{figure}
    \centering
\begin{tikzpicture}
\node at (-4.5,2.7) { $b$};
\draw (-4.5,2.7)  circle (0.25cm);
\node at (0,0) {\includegraphics[width=0.75\linewidth]{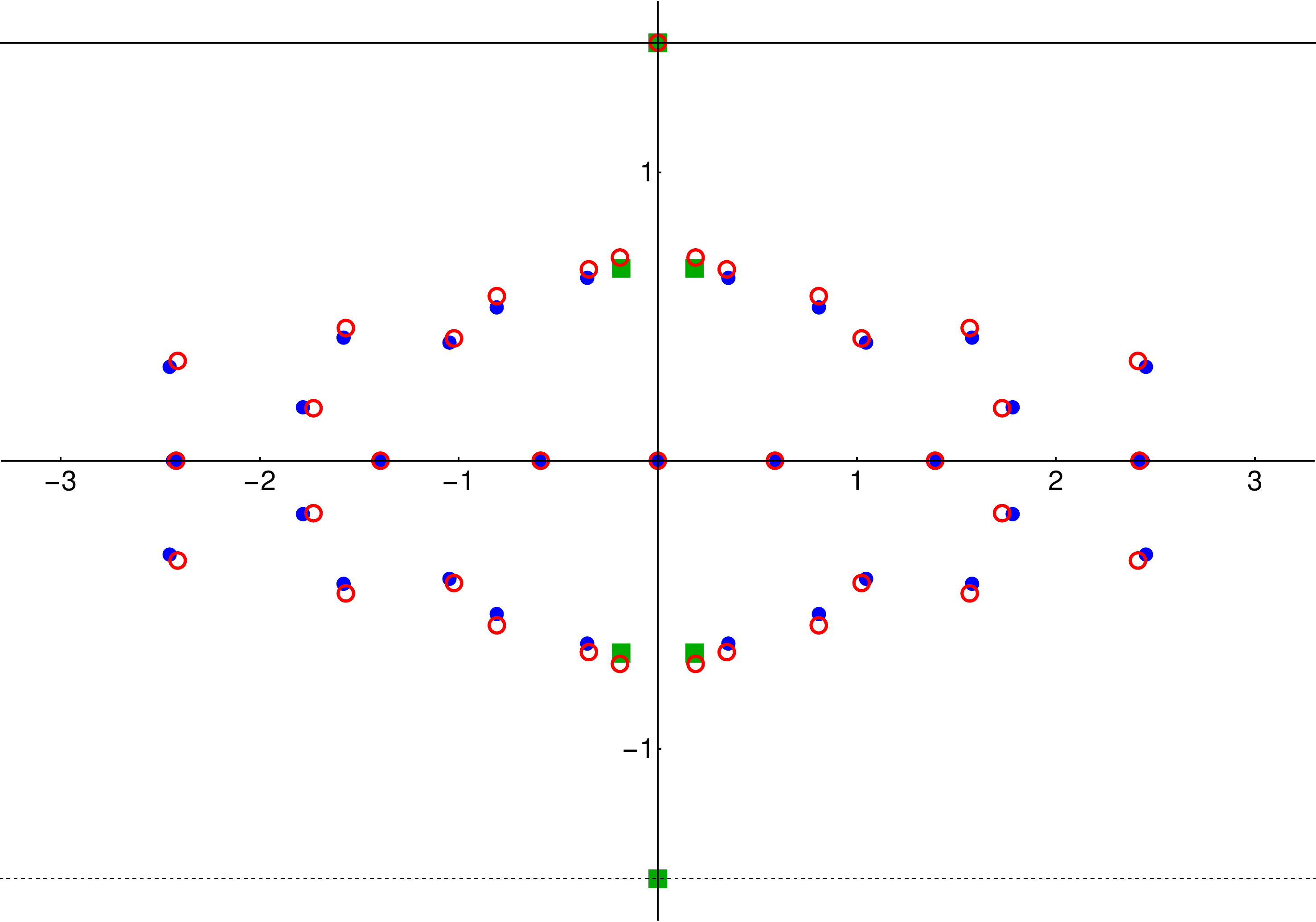}};
\node[anchor=west] at (3.65,4.0)   {$\Im m(b)=\frac{n}{2}$};
\node at (4.5,2.8) {$\texttt{d}=2$};
\node[anchor=west] at (3.5,-3.25)   {$\Im m(b)=-\frac{n}{2}$};
\end{tikzpicture}
\bigskip
\bigskip

\begin{tikzpicture}
\node at (-4.5,2.7) { $b$};
\draw (-4.5,2.7)  circle (0.25cm);
\node at (0,0) {\includegraphics[width=0.75\linewidth]{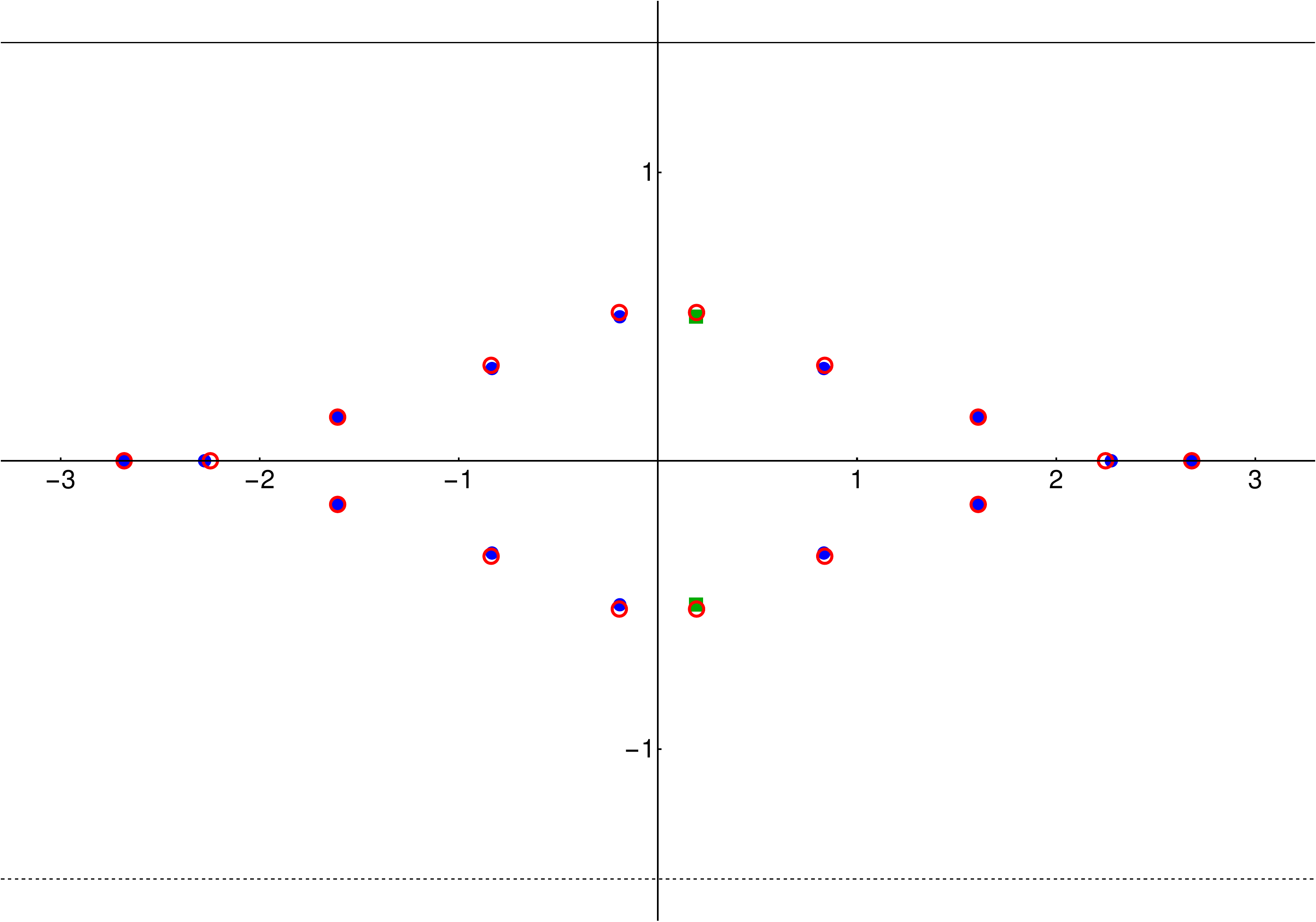}};
\node[anchor=west] at (3.65,4.0)   {$\Im m(b)=\frac{n}{2}$};
\node[anchor=west] at (3.5,-3.25)   {$\Im m(b)=-\frac{n}{2}$};
\node at (4.5,2.8) {$\texttt{d}=1$};
\end{tikzpicture}
    \caption{\label{sdsdjhsjdhsjd}\small Numerical data for $b(L)$ and $b_*(L)$ is plotted in the complex $b$ plane for
the lattice system with $L=10$. Out of the few hundred states that were considered, only those with $\texttt{d}=2$ (top panel),
$\texttt{d}=1$ (bottom panel) and $U_q\big(\mathfrak{sl}(2)\big)$ spin
${\cal S}=2$ were used to produce the figure. 
The open circles correspond to $b(L)$ that
was extracted from the Bethe roots by means of eqs.\,\eqref{ska823jhds},\,\eqref{ska823jhdsAAA}. 
The filled shapes  represent $b_*(L)$ obtained from an 
analysis of the quantization condition \eqref{kajs8712jhds1}. The green squares and blue circles are
used to distinguish whether in the scaling limit $b_*(L)$ becomes
a pure imaginary number or a real number, respectively. The two green squares in the top panel
for which $\slim_{L\to\infty}b_*(L)=\pm\frac{\ri n}{2}$ correspond to the same state. It seems interesting
to note that the agreement between $b(L)$ and $b_*(L)$ is better than in the case of the lattice model with
quasi-periodic boundary conditions imposed, compare the above figures with the ones
 contained in Appendix C  of ref.\cite{Bazhanov:2019xvyA}.
The parameter $n$ was taken to be $n=2.9$.}
    % State number is 28
    % Saved under spectro/RG/Z28
\end{figure}

A  verification of the quantization condition  \eqref{kajs8712jhds1} was carried out using numerical data
obtained from the lattice model
with $L=10$. The spin chain Hamiltonian was constructed 
and  the first few hundred lowest energy Bethe states
were found via a direct diagonalization procedure. Note that, because of the
$U_q\big(\mathfrak{sl}(2)\big)$ symmetry, it was sufficient to focus on the sector
with $S^z=0$ as there  is always one state $|\Psi_L\rangle$
 from the  $U_q\big(\mathfrak{sl}(2)\big)$  multiplet
${\cal M}_{\cal S}$ lying in this sector. For each Bethe state, apart from the energy, 
the eigenvalue of the quasi-shift operator was computed from which we extracted $b$.
The  numerical data for $b(L)$ was compared with  $b_*(L)$ --- the predictions coming
from the quantization condition. The latter was obtained by considering \eqref{kajs8712jhds1}
with $L=10$ and the correction term $O(L^{-\epsilon})$ ignored as an equation from which $b_*(L)$ 
could be  determined numerically. Note that the phase shift $\re^{\frac{\ri}{2}\delta}$  therein depends on $b$ transcendentally
via the Gamma functions as in  \eqref{askjjh1287sd} and algebraically
through the set $\pmb{w}$, which solves the coupled  system \eqref{sksksk1} with $s\mapsto b$.
For given  $\texttt{d}\le 3$ we took the ${\rm par}_2(\texttt{d})-{\rm par}_2(\texttt{d}-2{\cal S}-1)$
equations which are obtained from the quantization condition by specializing
the phase shift $\delta=\delta(\pmb{w},p,s)$ to different solution sets $\pmb{w}=\{w_a\}^\texttt{d}_{a=1}$ of \eqref{sksksk1}.
For each of them we found all possible solutions   $b_*(L)$
that  lie in a suitably chosen finite portion of the strip
$\big|\Im m(b_*)\big|<\frac{n}{2}+\varepsilon$ with $0<\varepsilon\ll 1$ of the complex $b$ plane.
Some of the results for the comparison of  $b(L)$  and $b_*(L)$ for $L=10$ are presented in
Fig.\,\ref{sdsdjhsjdhsjd}. They motivated us to make the following conjecture. 
\bigskip

\noindent
{\bf Conjecture:\label{MAIN_CONJ}}
For any RG trajectory $\{|\Psi_L\rangle\}$
labeled by ${\cal S}$, $\texttt{d}$
and a solution set $\pmb{w}=\{w_a\}_{a=1}^\texttt{d}$ of eq.\,\eqref{sksksk1}
 the corresponding value
of $b(L)=\frac{n}{2\pi}\log\big(\sqrt{B}\,\big)$,  with $\sqrt{B}$ computed according to 
formula \eqref{ska823jhds}, obeys the quantization condition 
\eqref{kajs8712jhds1}.
Conversely, let
$b_*(L)$ with    $-\frac{n}{2}<\Im m\big(b_*(L)\big)\le \frac{n}{2}$  be a solution of the relation   \eqref{kajs8712jhds1} with the correction
terms ignored. Then, there exists  a unique $U_q\big(\mathfrak{sl}(2)\big)$  multiplet
${\cal M}_{\cal S}$ for which $\sqrt{B}$ obtained from $|\Psi_L\rangle\in{\cal M}_{\cal S}$
is such that
$\sqrt{B}-\exp\big(\frac{2\pi}{n}\,b_*(L)\big)$ goes to zero  faster than any power of the logarithm of $L$.

\section{Space of states in the scaling limit \label{sec4}}

\subsection{Continuous and discrete spectrum\label{sec41}}

A key result of this paper is  the conjecture, above,
which was motivated by our numerical work.
It describes a  certain one-to-one relation between $b(L)$ and $b_*(L)$. 
The former is obtained via the Bethe roots 
corresponding to a state $|\Psi_L\rangle$ in a multiplet ${\cal M}_{\cal S}$, labeled
by $2p=2{\cal S}+1-(n+2)$,
the non-negative integer $\texttt{d}$ and one of the 
$N(\texttt{d}\,|\,{\cal S})={\rm par}_2(\texttt{d})-{\rm par}_2(\texttt{d}-2{\cal S}-1)$
solution sets $\pmb{w}=\{w_a\}_{a=1}^\texttt{d}$ of the algebraic system
 \eqref{sksksk1}. Since it requires the construction of an RG trajectory, computing $b(L)$
for $L\gg1$ can be cumbersome and time-consuming to carry out.
The notation $b_*(L)$ 
stands for a solution of the
 quantization condition \eqref{kajs8712jhds1}  treated as an equation for $b(L)$
with the correction terms ignored. 
Accepting the conjecture to be true, one can determine the spectrum of $b(L)$ for the low energy states at large $L$ 
via a study of \eqref{kajs8712jhds1}. This is extremely powerful, since  much less computing resources are needed
to numerically analyze the quantization condition and it may be studied analytically as well.
The results allow one to  characterize the spectrum of conformal dimensions together with the
space of low energy states in the scaling limit. 
\bigskip

Consider the quantization condition \eqref{kajs8712jhds1} and suppose that 
 $\delta(\pmb{w},p,s)|_{s\mapsto b(L)}\ll \log(L)$.
Then the first term dominates and one can develop  an asymptotic expansion for $b(L)$ in $1/\log(L)$. 
The leading and subleading asymptotic behaviour reads as 
\be\label{aksn12b3}
b_{\mathfrak{m}}(L)=\frac{2\pi\mathfrak{m}-\delta_0}{4\log\big(\re^{\frac{1}{4}\delta'_0}L/L_0\big)}+
O\big((\log L)^{-3}\big)
\qquad\qquad\qquad \big(L\gg 1,\ \mathfrak{m}\ {\rm-\ fixed}\big)\ .
\ee
Here we use the notation
\be
\delta_0=\delta|_{s=0}\,,\qquad\qquad\qquad \delta'_0=\partial_s\delta|_{s=0}\,,
\ee
while $\mathfrak{m}$, which labels the different $b(L)$ obeying  the quantization condition,
comes about as a result of taking the logarithm of  \eqref{kajs8712jhds1} and is an even or odd integer for
$\sigma=+1$ or $\sigma=-1$, respectively. Formula \eqref{aksn12b3} shows that, in general,
$b_{\mathfrak{m}}(L)$ is  a complex number  since $\delta_0$ and $\delta'_0$ are generically complex.
However, while $\Im m\big(b_{\mathfrak{m}}(L)\big)\sim1/\log(L)\to 0$ as $L\to\infty$
the magnitude of the real part is controlled
by the integer $\mathfrak{m}$ which, for the low energy states,
 may take any values up to some $\mathfrak{m}_{\rm max}\ll L$. Numerical
work leads us to suppose that
$\lim_{L\to\infty} b_{\mathfrak{m}_{\rm max}}= \infty$.
\medskip

Let ${\cal H}_{L|{\cal S}}^{({\rm cont})}$ denote the set of low energy states $|\Psi_L\rangle$
with fixed value of the $U_q\big(\mathfrak{sl}(2)\big)$
 spin ${\cal S}=0,1,2,\ldots$ such that $\Im m\big(b(L)\big)\to 0$ as $L\to\infty$.
Recall that the states come in multiplets ${\cal M}_{\cal S}$ and one can choose a basis
for that multiplet in which
the $z$\,-projection of the total spin, $\mathbb{S}^z=\frac{1}{2}\sum_{J=1}^{2L}\sigma_J^z$,
is diagonal.
This yields the refinement
\be
{\cal H}_{L|{\cal S}}^{({\rm cont})}=\bigcup_{S^z=-{\cal S}}^{{\cal S}} {\cal H}_{L|{\cal S},S^z}^{({\rm cont})}\ .
\ee
Each low energy Bethe state in ${\cal H}_{L|{\cal S},S^z}^{({\rm cont})}$
is uniquely specified by the non-negative integer $\texttt{d}$, a solution set $\pmb{w}=\{w_a\}_{a=1}^\texttt{d}$ 
of the algebraic system \eqref{sksksk1} with $s\mapsto b(L)$,
as well as the even or odd integer $\mathfrak{m}$ that enters into the asymptotics \eqref{aksn12b3}.
For $L\gg 1$ the value of $b_{\mathfrak{m}}(L)$ becomes densely distributed in the segment
$(-b_{\mathfrak{m}_{\rm max}}(L),+b_{\mathfrak{m}_{\rm max}}(L))$. The density of states is obtained
from the quantization condition \eqref{kajs8712jhds1} written in logarithmic form:
\be\label{ksajhiu3jds}
\big[4 s\log(L/L_0)+
\delta(\pmb{w},p,s)\big]|_{s\mapsto b_{\mathfrak{m}}(L)}=2\pi\mathfrak{m}+O(L^{-\epsilon})\ .
\ee
Here the branch of the logarithm needed to define $\delta$ from the relation \eqref{asnbbvb32vbAAAAv}
 is taken such that the phase shift is a continuous
function of $s$ in the strip $|\Im m(s)|<\varepsilon$ for some 
 $\varepsilon>0$ (it is being assumed that $\re^{\frac{\ri}{2}\delta}$
contains no zeroes or poles for real $s$). 
The term in the square brackets in the left hand side of \eqref{ksajhiu3jds} is a monotonic
function of $s$ for $L$ sufficiently large. This way one concludes that the number of states in 
$ {\cal H}_{L|{\cal S},S^z}^{({\rm cont})}$ with fixed $\texttt{d}$
such that $\Re e\big(b(L)\big)$ lies in the interval $(s,s+\Delta s)
\in(-b_{\mathfrak{m}_{\rm max}}(L),+b_{\mathfrak{m}_{\rm max}}(L))$ is given by $\rho^{(\texttt{d})}_p(s)\, \Delta s$  with\footnote{This line of arguments is analogous to the standard derivation in the root density approach. One introduces a monotonic increasing counting function which evaluates to (half-)integers at the Bethe roots similar as the l.h.s of \eqref{ksajhiu3jds} evaluates to odd/even integers $\mathfrak{m}$ multiplied by $2\pi$ when $s$ is swapped for $b_\mathfrak{m}(L)$
and $L\gg 1$. Differentiating the counting function in the root density approach yields the root density, while we obtain 
the density of states \eqref{kasnb12bdaaaa} by differentiating \eqref{ksajhiu3jds} and dividing by $4\pi$.}
\be\label{kasnb12bdaaaa}
\rho^{(\texttt{d})}_p(s)= \frac{1}{\pi}\,N(\texttt{d}\,|\,{\cal S})\,\log\big(2^{\frac{n+2}{n}}\,L/L_0\big)+
\frac{1}{2\pi\ri}\ \partial_s\log\Bigg[
\bigg(\frac{\Gamma(\frac{1}{2}+p-\ri s)}{\Gamma(\frac{1}{2}+p+\ri s)}\bigg)^{N(\texttt{d}\,|\,{\cal S})}\,
\prod_{\pmb{w}\atop \texttt{d}{\rm \,-\,fixed}}
\frac{\check{\mathfrak{C}}_{p,s}^{(+)}(\pmb{w})}{\check{\mathfrak{C}}^{(-)}_{p,s}(\pmb{w})}\,\Bigg]
\ee 
up to corrections which vanish as $L\to\infty$.
The product over $\pmb{w}$ appearing in the r.h.s. goes over all the
$N(\texttt{d}\, |\, \mathcal{S})$ \eqref{ksajnb389jhdsa} solution sets of the algebraic system \eqref{sksksk1}
with $\texttt{d}$ fixed.  Also, recall that $2p=2{\cal S}+1-(n+2)$. 
\medskip

In the work \cite{Kotousov:2019nvt} a formula is presented for a product over $\pmb{w}$
similar to the one appearing in the r.h.s. of \eqref{kasnb12bdaaaa} (see also Appendix B of \cite{Bazhanov:2019xvyA}). It is valid for the 
case of generic $p$ and $n$ when the number of solution sets $\pmb{w}$  of
 \eqref{sksksk1} is ${\rm par}_2(\texttt{d})$. Based on this, one can derive the result:
\bea\label{mn2398jsc}
\prod_{\pmb{w}\atop \texttt{d}{\rm \,-\,fixed}}
\frac{\check{\mathfrak{C}}_{p,s}^{(+)}(\pmb{w})}{\check{\mathfrak{C}}^{(-)}_{p,s}(\pmb{w})}&=&
(-1)^{{\rm par}_2(\texttt{d}-2{\cal S}-1)}
\prod_{a=0}^{\texttt{d}-1}\bigg(\frac{\frac{1}{2}+a+p-\ri s}{\frac{1}{2}+a+p+\ri s}\bigg)^{%
N(\texttt{d}\,|\,{\cal S})-N_a^+(\texttt{d}\,|\,{\cal S})}\nonumber\\[0.2cm]
&\times&
\prod_{a=0}^{\texttt{d}-1}\bigg(\frac{\frac{1}{2}+a-p-\ri s}{\frac{1}{2}+a-p+\ri s}\bigg)^{%
N(\texttt{d}\,|\,{\cal S})-N_a^-(\texttt{d}\,|\,{\cal S})}
\eea
with the integers $N^\pm_a$ being defined through their generating function as
\be\label{askjb32bvbvsd}
\sum_{\texttt{d}=0}^\infty N_a^\pm(\texttt{d}\,|\,{\cal S})\, \texttt{q}^\texttt{d}=
\bigg(\prod_{j=1}^\infty (1-\texttt{q}^j)^{-2}\bigg)\,\sum_{m=0}^\infty (-1)^m\,
\big(1-\texttt{q}^{(1\pm m)(2{\cal S}+1)}\big)\,\texttt{q}^{ma+\frac{m(m+1)}{2}}\ .
\ee
Notice that
\be
N_a^+(\texttt{d}\,|\,{\cal S})=N(\texttt{d}\,|\,{\cal S})-N_{-a-1}^-(\texttt{d}\,|\,{\cal S})\, .
\ee
\medskip

The scaling limit of the
RG trajectory $L\mapsto|\Psi_L\rangle\in{\cal H}_{L|{\cal S}}^{({\rm cont})}$ labeled by real $s$,
the integers ${\cal S}$, $S^z$, $\texttt{d}$ and the solution set $\pmb{w}$ yields
\be\label{asmnnb3298as}
\slim_{L\to\infty} |\Psi_L\rangle=|\psi_{p,s}^{(S^z)}(\pmb{w})\rangle\ .
\ee
One can define the linear span ${\cal H}_{{\cal S}}^{({\rm cont})}$ 
of all such possible states with fixed ${\cal S}$. The above discussion implies that this linear space
admits the decomposition
\be\label{sakj9832jh}%
{\cal H}_{\cal S}^{({\rm cont})} =\bigoplus_{S^z=-{\cal S}}^{{\cal S}}
{\cal H}_{{\cal S},S^z}^{({\rm cont})} \,,
\ee
where each of the spaces ${\cal H}_{{\cal S},S^z}^{({\rm cont})}$ is isomorphic to
${\cal H}_{{\cal S},{\cal S}}^{({\rm cont})}$ and
\be\label{kasnb12bvad}
{\cal H}_{{\cal S},{\cal S}}^{({\rm cont})}=\int^{\oplus}_{\mathbb{R}}\rd s\ 
\bigoplus_{\texttt{d}=0}^\infty {\cal H}^{({\rm cont},\texttt{d})}_{p,s}\qquad\qquad\qquad
\qquad\qquad \big(2p=2{\cal S}+1-n-2\big)\, .
\ee
The components appearing inside the direct sum are finite dimensional such that
\be\label{kasnb12bvadBBBB}
\dim\big({\cal H}_{p,s}^{({\rm cont},\texttt{d})}\big)=N(\texttt{d}\,|\, \mathcal{S})\ .
\ee
\medskip

For the low energy states 
where the value of $\Im m\big(b(L)\big)$ is nonvanishing in the limit $L\to\infty$
so that they do not belong to  ${\cal H}_{L|{\cal S}}^{({\rm cont})}$, one may repeat the similar
analysis that was performed in ref.\cite{Bazhanov:2019xvyA} in the context of 
the staggered six-vertex model with quasi-periodic
BCs (see Appendix B therein). Let's denote by ${\cal H}_{L|{\cal S},S^z}^{({\rm disc})}$
the set of such states $|\Psi_L\rangle$ with given quantum numbers ${\cal S}$ and $S^z$.
The quantization condition \eqref{kajs8712jhds1}  implies that
the set $\pmb{w}$ and $s=\slim_{L\to\infty} b(L)$ labelling the RG trajectory $\{|\Psi_L\rangle\}$ must be such that
\be\label{askjbnbv21984}
\re^{-\frac{\ri}{2}\delta(\pmb{w},p,s)}=0\qquad {\rm if }\qquad \Im m(s)>0\,,\qquad\qquad\qquad
\re^{+\frac{\ri}{2}\delta(\pmb{w},p,s)}=0
\qquad {\rm if }\qquad \Im m(s)<0\, .
\ee 
We supplement this with the additional constraint
on the imaginary part on $s$:
\be\label{askn32bn89}
-\frac{n}{2}<\Im m(s)\le \frac{n}{2}\,.
\ee
It comes from the inequality \eqref{ska823jhdsBBB}, while
the line $\Im m(s)=-\frac{\ri n}{2}$ was excluded from the interval in order to avoid double counting states with $\slim_{L\to\infty} b(L)=\pm\frac{\ri n}{2}$.
It turns out that the phase shift satisfies:
\be\label{mnaaaaas8912n}
\re^{\frac{\ri}{2}\delta(\pmb{w},p,s)}=\re^{-\frac{\ri}{2}\delta(-\pmb{w},p,-s)}\,.
\ee
Here $-\pmb{w}$
denotes the set $\{-w_a\}_{a=1}^\texttt{d}$, where if $\pmb{w}$ obeys
the algebraic system
\eqref{sksksk1}  then $-\pmb{w}$ obeys the same set of equations with $s\mapsto-s$.
This allows one to focus on the case with
$0<\Im m(s)\le \frac{n}{2}$,
while results for $-\frac{n}{2}<\Im m(s)<0$ follow by simply flipping the sign $s\mapsto-s$.
\medskip

The analysis of
\eqref{askjbnbv21984} is greatly facilitated by
the relation
\be\label{ajsnb218874}
\prod_{\pmb{w}\atop \texttt{d}{\rm \,-\,fixed}}\re^{\frac{\ri}{2}\delta(\pmb{w},p,s)}=
\bigg(2^{(2n+4)\frac{\ri s}{n}}\ \frac{\Gamma(\frac{1}{2}+p-\ri s)}{\Gamma(\frac{1}{2}+p+\ri s)}\bigg)^{N(\texttt{d}\,|\,{\cal S})}\,
\prod_{\pmb{w}\atop \texttt{d}{\rm \,-\,fixed}}
\frac{\check{\mathfrak{C}}_{p,s}^{(+)}(\pmb{w})}{\check{\mathfrak{C}}^{(-)}_{p,s}(\pmb{w})}
\ee 
with the last term in the r.h.s. being given by the product \eqref{mn2398jsc}. 
It follows  from the definitions
\eqref{asnbbvb32vbAAAAv},\,\eqref{nsanb3vb23AAA} and \eqref{nsanb3vb23BBB}.
Also, we'll need the following assumptions on the positions of the poles and zeroes
 of the function $\re^{\frac{\ri}{2}\delta(\pmb{w},p,s)}$, which
were verified numerically for small $\texttt{d}\le 3$:
\begin{itemize}
\item[(i)] The points where
 $\re^{\frac{\ri}{2}\delta(\pmb{w},p,s)}$ is singular do not coincide with the location of any zero
of $\re^{\frac{\ri}{2}\delta(\pmb{w}',p,s)}$ with $\pmb{w}'$ being some other
solution set of \eqref{sksksk1}.
\item[(ii)] 
All singularities of $\re^{\frac{\ri}{2}\delta(\pmb{w},p,s)}$ in the complex $s$ plane are simple poles.
Notice that, in view of eq.\,\eqref{mnaaaaas8912n}, this implies 
that all of its zeroes are simple as well.
\end{itemize}
From assumption (i), any 
 pole or zero of  $\re^{\frac{\ri}{2}\delta(\pmb{w},p,s)}$ must appear as a pole/zero in the r.h.s.
of  \eqref{ajsnb218874}. This way, one finds 
that the values of $s$ for which the first condition in 
\eqref{askjbnbv21984} is obeyed are $s=\pm s_a$, with
\be\label{kjasnb1221sa}
s_a=\ri\big(-p-\tfrac{1}{2}-a\big)=
\ri\big(\tfrac{n}{2}-{\cal S}-a\big)\qquad\qquad {\rm and}\qquad\qquad 0\le a+{\cal S}< \tfrac{n}{2}\,, \ 
\ \ a\in\mathbb{Z}\, ,
\ee
where the bound on $a+{\cal S}$ comes from \eqref{askn32bn89}.
Moreover, due to (ii) one can determine the number of solution sets $\pmb{w}$ with
\eqref{askjbnbv21984} being satisfied at $s=s_a$ by reading off the multiplicity of that
pole/zero from eqs.\,\eqref{ajsnb218874} and \eqref{mn2398jsc}. 
 This  would coincide with
the dimension of the linear space ${\cal H}_{p,s}^{({\rm disc},\texttt{d})}$,
which is the span of all states of the form $|\psi_{p,s}^{(S^z)}(\pmb{w})\rangle$
having fixed ${\cal S}$,
$S^z$, $\texttt{d}$ and $s$  with $\Im m(s)\ne0$.
One finds the number of such $\pmb{w}$
to be $N_a^+(\texttt{d})$.

\medskip

Define the space ${\cal H}_{{\cal S},S^z}^{({\rm disc})}$ as the linear span 
of all the states that appear in the scaling limit of ${\cal H}_{L|{\cal S},S^z}^{({\rm disc})}$.
These  are isomorphic to ${\cal H}_{{\cal S},{\cal S}}^{({\rm disc})}$and 
the analysis of the quantization condition performed above implies that:
\be\label{samnnb123AAA}
{\cal H}_{{\cal S},{\cal S}}^{({\rm disc})}=\bigoplus_{s\in\Sigma^+\cup\Sigma^-}\ 
\bigoplus_{\texttt{d}=0}^\infty
{\cal H}_{p, s}^{({\rm disc},\texttt{d})}\ .
\ee
Here $\Sigma^\pm$ denote the finite sets of pure imaginary numbers:
\be\label{328daaa7ssa}
\Sigma^+=\big\{s\,:\ \tfrac{n}{2}+\ri s\in\mathbb{Z} \,,\ \Im m(s)\in(0,\tfrac{n}{2}]\big\}\,,\quad
\qquad 
\Sigma^-=\big\{s\,:\ \tfrac{n}{2}-\ri s\in\mathbb{Z} \,,\ \Im m(s)\in(-\tfrac{n}{2},0)\big\}\,,
\ee
which incorporate the bound on the imaginary part of $s$ \eqref{askn32bn89}.
Each component ${\cal H}_{p,s}^{(\texttt{d})}$ is finite dimensional and
\be\label{ask98aaa32jd}
\dim\big({\cal H}_{p,s}^{({\rm disc},\texttt{d})}\big)=N^+_a(\texttt{d}\,|\,{\cal S})\qquad
\qquad {\rm with}\qquad\qquad
 a=\tfrac{n}{2}-{\cal S}\pm\ri s\in\mathbb{Z}
\ee
(here and below, when a condition involving $\pm \ri s$ appears we mean it is to be satisfied for some choice of the sign $+$ or $-$). 
\medskip

The following comment is in order here. For the case $a<0$, the integers $N_a^+(\texttt{d}\,|\,{\cal S})$ 
\eqref{askjb32bvbvsd} are all zero for $\texttt{d}=0$:
\be
N_a^+(0\,|\,{\cal S})=0\qquad{\rm for}\qquad a=-1,-2,-3,\ldots\ .
\ee
As a result, for the primary Bethe states the limiting values of $\Im m\big(b(L)\big)$ are 
given by $s=\pm s_a$ \eqref{kjasnb1221sa} with the extra condition imposed that $a\ge 0$.
Thus one recovers the results of the work \cite{Frahm:2021ohj}, see also formula \eqref{askhbaaa23}.
Of course, RG trajectories exist for the spin chain which are not primary Bethe states that
 are labeled by $s=\pm s_a$ with $a<0$.
\medskip

We conjecture that any RG trajectory of the lattice model with given $U_q\big(\mathfrak{sl}(2)\big)$
spin ${\cal S}$ and eigenvalue of the $z$\,-projection of the total
spin operator $S^z$ belongs either to ${\cal H}_{L|{\cal S},S^z}^{({\rm cont})}$ or ${\cal H}_{L|{\cal S},S^z}^{({\rm disc})}$.
Thus, the full space of low energy states of the lattice system in the scaling limit becomes the linear space
\be\label{mn3287sdjh}
{\cal H}={\cal H}^{({\rm cont})}\oplus {\cal H}^{({\rm disc})}
\ee
with
\be\label{samnnb123}
{\cal H}^{({\rm cont})}=\bigoplus_{{\cal S}=0}^{\infty}\,\bigoplus_{S^z=-{\cal S}}^{{\cal S}}
{\cal H}_{{\cal S},S^z}^{({\rm cont})}\,,\qquad\qquad
{\cal H}^{({\rm disc})}=\bigoplus_{{\cal S}=0}^{\infty}\,\bigoplus_{S^z=-{\cal S}}^{{\cal S}}
{\cal H}_{{\cal S},S^z}^{({\rm disc})}\ .
\ee
We call ${\cal H}^{({\rm cont})}$
the `continuous spectrum' due to the presence of a direct integral over $s$ in its decomposition, see \eqref{kasnb12bvad}. The space ${\cal H}^{({\rm disc})}$ will be referred to as the `discrete spectrum'.

\subsection{\texorpdfstring{${\cal W}_\infty$}{W} algebra}
In the scaling limit the critical lattice system possesses extended conformal symmetry.
The corresponding algebra is expected to be the ${\cal W}_\infty$ algebra from ref.\cite{Bakas:1991fs}
with central charge $c<2$. This is
the same one that appears in the scaling limit of the staggered
six-vertex model with (quasi-)periodic BCs \cite{Bazhanov:2019xvy,Bazhanov:2019xvyA}.
Among other things, such a statement implies that  the graded linear spaces
\be\label{nnb32vb2121}
\bigoplus_{\texttt{d}=0}^\infty{\cal H}_{p,s}^{(\texttt{d})}\,,\qquad\qquad
{\cal H}_{p,s}^{(\texttt{d})}=\begin{cases}
{\cal H}_{p,s}^{({\rm  cont},\texttt{d})}& \ \ \ {\rm for}\qquad s\in\mathbb{R}\\[0.2cm]
{\cal H}_{p,s}^{({\rm  disc},\texttt{d})}& \ \ \ {\rm for}\qquad p+\tfrac{1}{2}\pm\ri s\in\mathbb{Z}
\end{cases}
\ee
are isomorphic to a (irreducible) representation of ${\cal W}_\infty$. Then formulae \eqref{samnnb123},\,
\eqref{kasnb12bvad} and \eqref{samnnb123AAA} would provide a classification of the space
of states ${\cal H}$ occuring in the scaling limit of the lattice model 
 in terms of irreps of the algebra of extended conformal symmetry. 
In order to demonstrate this we briefly mention some details concerning the ${\cal W}_\infty$
algebra and its representations, while referring the reader to section 16 of ref.\cite{Bazhanov:2019xvyA}
 for a deeper discussion.
\medskip

The ${\cal W}_\infty$ algebra is generated by a set of currents $W_j(u)$
of Lorentz spin $j=2,3,\ldots\ $.
These satisfy an infinite system of Operator Product Expansions (OPEs).
Its first few members can be chosen to be
 \bea\label{aiisaisa}
 W_2(u)\,W_2(0)&=&\frac{c}{2 u^4}-\frac{2}{u^2}\ W_2(0)-\frac{1}{u}\ \partial W_2(0)+O(1)
 \nonumber \\[0.2cm]
W_2(u)\,W_3(0)&=&-\frac{3}{u^2}\ W_3(0)-\frac{1}{u}\ \partial W_3(0)+O(1)
\\[0.2cm]
W_3(u)\,W_3(0)&=&-\frac{c(c+7)(2c-1) }{9(c-2)u^6}+\frac{ (c+7)(2c-1)}{3(c-2)u^4}\ 
\big(W_2(u)+W_2(0)\big)-\frac{1}{u^2}\ \Big( W_4(u)+W_4(0)\nonumber \\[0.2cm]
&+&W^2_2(u)+W^2_2(0)
+\frac{2c^2+22c-25}{30 (c-2)}\,\big (\partial^2W_2(u)+\partial^2W_2(0)\big)\Big)+O(1)\ ,
\nonumber
\eea
where in the last line $W_2^2$ is a composite field which coincides with
the first regular term in the OPE $W_2(u)W_2(0)$. 
Notice that there is some ambiguity in the definition of $W_j$
for $j\ge 3$.
Apart from the freedom in the overall multiplicative normalization, $W_j\mapsto C W_j$,
it is possible to add to $W_j$ any differential polynomial of Lorentz spin $j$ involving the lower spin currents $W_k$
with $k<j$. Here, the $W_3$ current
was fixed by the requirement that it be a primary field of spin three, so that its OPE
with $W_2$ takes the form of the second line in formula \eqref{aiisaisa}. As for $W_4$, one
 can not  arrange for it to be a primary field by adding linear combinations of $W_2^2$,
$\partial W_3$ and $\partial^2 W_2$. Defined such that it appears in the OPE of
$W_3(u)W_3(0)$ as above, it turns out that
$W_2(u)W_4(0)$ takes a simpler form,
\be\label{8989aioiaso}
 W_2(u)\,W_4(0)=\frac{(c+10)(17c+2)}{15 (c-2)\, u^4}
 \ W_2(0)-\frac{4}{u^2}\ W_4(0)-\frac{1}{u}\ \partial W_4(0)+O(1)\ ,
\ee
 where the 
singular terms $\propto u^{-6}$ and $\propto u^{-3}$ are absent.
\medskip

For the study of the ${\cal W}_\infty$ algebra it is useful to know that it
admits a realization in terms of two independent
chiral Bose fields. We normalize them as
\be
\partial\vartheta(u)\,\partial\vartheta(0)=-\frac{1}{2u^2}+O(1)\,,\qquad\qquad
\partial\varphi(u)\,\partial\varphi(0)=-\frac{1}{2u^2}+O(1)\,,
\ee
while $\partial\varphi(u)\partial\vartheta(0)=O(1)$. One may check that as a consequence of the free field OPEs,
the currents
\bea\label{w2iosdi}
W_2&=&(\partial\vartheta)^2+(\partial \varphi)^2+\frac{\ri}{\sqrt{n+2}}\ \partial^2\varphi\\[0.2cm]
W_3&=&
\frac{6n+8}{3n+6}\, (\partial \vartheta)^3+2\,
 (\partial \varphi)^2\partial \vartheta+\ri\sqrt{n+2}\ \partial^2 \varphi\,\partial\vartheta
-\frac{\ri n}{\sqrt{n+2}}\ \partial\varphi\,\partial^2\vartheta+\frac{n}{6(n+2)}\  \partial^3\vartheta \nonumber
\eea
obey the algebra \eqref{aiisaisa}. The parameter $n$ entering above is related to the
central charge $c$ as
\be
c=\frac{2(n-1)}{n+2}
\ee
so that if $n$ is real and positive, the central charge $c$ is less than two.
Notice that, while an expression for $W_4$ in terms of $\partial\vartheta$ and $\partial\varphi$
has not been provided, it can be deduced from the OPEs \eqref{aiisaisa}
and the formula \eqref{w2iosdi} for $W_2$ and $W_3$.  One simply computes $W_3(u)W_3(0)$
with $W_3$ written
in terms of free fields and compares the coefficient $\propto u^{-2}$ with the same coefficient
appearing in the last two lines of eq.\,\eqref{aiisaisa}. It turns out that the higher spin currents
always appear in the OPEs involving the lower spin ones. This way, starting from 
\eqref{w2iosdi} and recursively computing OPEs, one can determine the
realization of   $W_j$ in terms of the free fields $\partial\varphi$ and $\partial\vartheta$
for any $j= 4,5,6,\ldots\ $. 
\medskip

A stepping stone for the construction of highest weight irreducible representations of the ${\cal W}_\infty$ algebra
is the Verma module. It is defined using the Fourier modes
of $W_j(u)$, which we assume to be periodic functions of the variable $u\sim u+2\pi$:
 \be\label{aoisd182918}
 W_{j}=-\frac{c}{24}\,\delta_{j,2}+\sum_{m=-\infty}^\infty {\widetilde W}_j(m)\ \re^{-\ri m u}  \ .
 \ee
Introduce the highest state, which is specified by the conditions:
\be\label{aios3298}
\widetilde{W}_j(m)\,|\pmb{\omega}\rangle=0 \qquad\qquad (\forall m>0)\,,\qquad\qquad
\widetilde{W}_j(0)\,|\pmb{\omega}\rangle=\omega_j\,|\pmb{\omega}\rangle
\ee
with $j=2,3$.
 The highest weight is given by $\pmb{\omega}=(\omega_2,\omega_3)$, where
the component $\omega_2$ is equal to the conformal dimension of the highest state, while
$\omega_3$ is the eigenvalue of $\widetilde{W}_3(0)$, which commutes with
$\widetilde{W}_2(0)$. The Verma module is spanned by the states that are obtained
by acting with the `creation modes' of the spin 2 and spin 3 currents on the highest state:
\be
\widetilde{W}_{2}(-\ell_1)\ldots\widetilde{W}_{2}(-\ell_m)\,
\widetilde{W}_{3}(-\ell_{1}')\ldots\widetilde{W}_{3}(-\ell_{m'}')|\pmb{\omega}\rangle
\ee
with $\ell_j,\,\ell'_{j'}\ge 1$.
It possesses a natural grading given by
\be
\texttt{d}=\sum_{j=1}^m \ell_j+\sum_{j=1}^{m'}\ell_j'
\ee
and the dimensions of its level subspace with fixed $\texttt{d}$ is  the number of bi-partitions of $\texttt{d}$, i.e.,
${\rm par}_2(\texttt{d})$ \eqref{mas873hsd}.
In what follows we will parameterize the highest weight for the Verma module
${\cal V}_{\rho,\nu}$ as
\bea\label{Deltvarpi1a}
\omega_2&=&\frac{\rho^2-\frac{1}{4}}{n+2}+\frac{\nu^2}{n}  \\[0.2cm]
\omega_3&=&
\frac{2\nu}{\sqrt{n}}\,\Big(\,\frac{\rho^2}{n+2}+\frac{(3n+4)\,\nu^2}{3n\,(n+2)}-\frac{2n+3}{12\,(n+2)}\,\Big)\nonumber
\ .
\eea
This is motivated by the free field realization \eqref{w2iosdi}. Supposing that the 
highest state is an eigenvector of the
operators $\int\rd u \,\partial\vartheta(u)$ and  $\int\rd u\,\partial\varphi(u)$ with eigenvalues
$\frac{\nu}{\sqrt{n}}$ and $\frac{\rho}{\sqrt{n+2}}$, respectively,  formula
\eqref{Deltvarpi1a} follows from
\eqref{w2iosdi}.
The highest weight is an even function of $\rho$.
As a result the spaces ${\cal V}_{\rho,\nu}$ and ${\cal V}_{-\rho,\nu}$ should be identified.
In the parameterization \eqref{Deltvarpi1a}, the conformal dimensions of a state in the Verma module at level $\texttt{d}$ is
such that
\be
\widetilde{W}_2(0)-\frac{c}{24}=\frac{\rho^2}{n+2}+\frac{\nu^2}{n}-\frac{1}{12}+\texttt{d}\, ,
\ee
which should be compared with eq.\,\eqref{sa8721hdsb}.
\medskip

For generic complex values of $\rho$ and $\nu$ the Verma module ${\cal V}_{\rho,\nu}$ is an irreducible representation
of the ${\cal W}_\infty$ algebra. However, with $\rho$, $\nu$ obeying certain constraints, the Verma module
contains null vectors -- highest states occurring at non-zero levels. Then the highest weight irrep ${\cal W}_{\rho,\nu}$
 can be obtained 
from ${\cal V}_{\rho,\nu}$ by factoring out all of the invariant subspace(s) generated by the null vector(s).
In view of applications to the scaling limit of the lattice model of particular interest is when
$\rho=\pm\frac{1}{2}\big(r-m(n+2)\big)$ with $r,m=1,2,\ldots\ $. In this case
a null vector occurs at level $\texttt{d}=mr$ and the Verma module splits into the direct sum
of two representations, which are irreducible for generic $n$ and $\nu$:
\be
{\cal V}_{\rho,\nu}= {\cal W}_{\rho,\nu}\oplus {\cal W}_{\rho',\nu}\qquad\qquad
{\rm with}\qquad \qquad
\begin{array}{ll}
\rho\phantom{'}=\frac{1}{2}\,(r-m\,(n+2))\qquad\qquad
(n,\,\nu\,{\rm -\,generic})\\[0.2cm]
\rho'=\frac{1}{2}\,(r+m\,(n+2))\qquad\qquad (r,m=1,2,\ldots)
\end{array}\, .
\ee
The space ${\cal W}_{\rho',\nu}$ is isomorphic to the Verma module and the dimensions
of its level subspaces is ${\rm par}_2(\texttt{d})$, while for ${\cal W}_{\rho,\nu}$, 
the level subspaces are ${\rm par}_2(\texttt{d})-{\rm par}_2(\texttt{d}-mr)$ dimensional. 
Consider again the components ${\cal H}^{({\rm cont},\texttt{d})}_{p,s}$, which appear in 
the decomposition of the continuous spectrum of the space of states occurring in the scaling limit
of the spin chain. Taking into account formulae \eqref{kasnb12bvadBBBB} and \eqref{ksajnb389jhdsa} it is clear that
\be
{\cal W}_{p,s}\cong\bigoplus_{\texttt{d}=0}^\infty {\cal H}^{({\rm cont},\texttt{d})}_{p,s}
\qquad\qquad\qquad \big(2p=2{\cal S}+1-(n+2)\,,\ s\ {\rm -\,real}\big)\, .
\ee
\medskip

To describe the discrete spectrum in terms of irreps of the ${\cal W}_\infty$ algebra, 
it is necessary to analyze the case when $\nu$ is such that  
$\rho+\frac{1}{2}\pm\ri\nu$ is an integer for some choice of the sign $\pm$.
As explained in, e.g., section 16.2 of reference \cite{Bazhanov:2019xvyA} 
the Verma module
with $\rho+\frac{1}{2}+\ri\nu=-a_+=0,\pm 1,\pm2,\ldots$ 
contains a null vector at level $|a_++\frac{1}{2}|+\frac{1}{2}$,
while for $-\rho+\frac{1}{2}+\ri\nu=-a_-=0,\pm 1,\pm2,\ldots$ 
there is a null vector at level $|a_-+\frac{1}{2}|+\frac{1}{2}$. Assuming $\rho$ is generic for now
the character of the irreducible representation,
\be\label{chdef1a}
{\rm ch}_{\rho,\nu}(\texttt{q})\equiv{\rm Tr}_{{\cal W}_{\rho,\nu}}\Big[\,\texttt{q}^{\widetilde{W}_2(0)-\frac{c}{24}}\,\Big]\,,
\ee
 is given by  \cite{Jayaraman:1989tu} (see also \cite{Gerasimov:1989mz,Griffin:1990fg}) 
\be\label{charadegen1a}
{\rm ch}_{\rho,\nu}(\texttt{q})=\texttt{q}^{-\frac{1}{12}+\frac{\nu^2}{n}+\frac{\rho^2}{n+2}}
\ \bigg(\prod_{m=1}^{\infty}\frac{1}{(1-\texttt{q}^m)^{2}}\bigg)
\sum_{m=0}^\infty (-1)^{m}\ \texttt{q}^{m|a+\frac{1}{2}|+\frac{m^2}{2}}\qquad\quad
\begin{array}{l}
\rho+\tfrac{1}{2}\pm\ri\nu=-a\in\mathbb{Z}\  \\[0.2cm]
n,\,\rho\ \ {\rm generic}
\end{array}\,.
\ee
If, in addition to $\nu$ being constrained as above,
$\rho\to \frac{1}{2}\,(2{\cal S}-n-1)$ then the irrep ${\cal W}_{\rho,\nu}$
further breaks up into two irreducible representations. One of them is generated by the
null vector which appears at level $2{\cal S}+1$ and has highest weights
$(\rho',\nu)$ with $\rho'=\frac{1}{2}\,(2{\cal S}+n+3)$.  Its character is given by
\eqref{charadegen1a} with $\rho$ replaced by $\rho'$ and 
$a\mapsto a'=-2({\cal S}+1)-a$. Taking the difference 
${\rm ch}_{\rho,\nu}(\texttt{q})-{\rm ch}_{\rho',\nu}(\texttt{q})$ with 
$\rho\to \frac{1}{2}\,(2{\cal S}-n-1)$ and $\rho'\to \frac{1}{2}\,(2{\cal S}+n+3)$
yields for
 the character of the irreducible representation ${\cal W}_{\rho,\nu}$  with
\be
\rho+\tfrac{1}{2}\pm\ri\nu=-a\in\mathbb{Z}\qquad\qquad {\rm and}\qquad\qquad 2\rho= 2{\cal S}-n-1
\ee
that
\be
{\rm ch}_{\rho,\nu}=\texttt{q}^{-\frac{1}{12}+\frac{\nu^2}{n}+\frac{\rho^2}{n+2}}
\ \bigg(\prod_{j=1}^{\infty}\frac{1}{(1-\texttt{q}^j)^{2}}\bigg)\sum_{m=0}^\infty(-1)^m\,q^{\frac{m^2}{2}}\,
\Big(q^{m|a+\frac{1}{2}|}- q^{2{\cal S}+1+m|2{\cal S}+a+\frac{3}{2}|}\Big)\ .
\ee
For the case $a\ge0$ the above expression, apart from the overall factor
 $\texttt{q}^{-\frac{1}{12}+\frac{\nu^2}{n}+\frac{\rho^2}{n+2}}$, coincides with the generating function 
\eqref{askjb32bvbvsd}
for the integers $N_a^+(\texttt{d}\,|\,{\cal S})$. This way, one concludes
\be
{\cal W}_{p,s}\cong\bigoplus_{\texttt{d}=0}^\infty {\cal H}^{({\rm disc},\texttt{d})}_{p,s}
\qquad\qquad \big(p={\cal S}+\tfrac{1}{2}-\tfrac{1}{2}(n+2)\,,\ 
 \tfrac{n}{2}-{\cal S}\pm\ri s=a\in\mathbb{Z}_{\ge 0}\big)\, .
\ee
\medskip

The remaining case to be considered is when $-\mathcal{S}\le a<0$. The lower bound comes from
the condition $s\in(-\frac{n}{2},\frac{n}{2}]$ 
which implies  that $\pm\ri s=\tfrac{n}{2}-{\mathcal S}-a\le \frac{n}{2}$.  From the definition of the integers
$N^+_a(\texttt{d}\,|\,{\cal S})$ \eqref{askjb32bvbvsd}, which give the dimensions of the level subspaces 
${\cal H}_{p,s}^{({\rm disc},\texttt{d})}\subset {\cal H}^{({\rm disc})}$, one finds
\be
\dim\big({\cal H}_{p,s}^{(\texttt{d})}\big)=0\qquad {\rm for} \qquad \texttt{d}=0,1,\ldots, |a|-1
\qquad\qquad \big(-p-\tfrac{1}{2}\pm\ri s=a\in\mathbb{Z}\,,\  -{\cal S}\le a <0\big)
\ee
Thus the corresponding irrep \eqref{nnb32vb2121} has highest state whose conformal
dimension is given by:
\be
\Delta=\frac{p^2}{n+2}+\frac{s^2}{n}+|a|\qquad\qquad \big(-\mathcal{S}\le a<0\big)\, .
\ee
This turns out to be an irreducible representation of the ${\cal W}_\infty$ algebra,
\be
{\cal W}_{\rho,\nu}=\bigoplus_{\texttt{d}=0}^\infty{\cal H}_{p,s}^{(\texttt{d})}\,,
\ee
with highest weight
parameterized as in \eqref{Deltvarpi1a}, where 
\be
\rho=\mathcal{S}+\frac{1}{2}\,,\qquad\quad \nu=\begin{cases} s-\frac{\ri n}{2} & {\rm for} \ 
\ \ (-\ri s) >0\\[0.2cm]
 s+\frac{\ri n}{2} & {\rm for} \ 
\ \ (-\ri s) <0\end{cases}\qquad\quad 
 \big( \tfrac{n}{2}-{\cal S}\pm\ri s=a\in\mathbb{Z}_{<0}\,,\  -{\cal S}\le a <0\big)\, .
\nonumber
\ee
Assuming $n$ is irrational, the character of such a representation is given by
\be\label{charadegen1v}
{\rm ch}_{\rho,\nu}(\texttt{q})=\texttt{q}^{-\frac{1}{12}+\frac{\nu^2}{n}+\frac{\rho^2}{n+2}}
 \  \bigg(\prod_{m=1}^{\infty}\frac{1}{(1-\texttt{q}^m)^{2}}\bigg)\ 
\sum_{m=0}^\infty (-1)^{m}\, \texttt{q}^{\frac{m^2}{2}}\big(\, \texttt{q}^{m |\, |\rho|-|\nu|\, |}-\texttt{q}^{(m+1)
(|\rho|+|\nu| +1 )-\frac{1}{2}}\,\big)\, .
\ee
One can check that the dimensions of the level subspaces, obtained by expanding ${\rm ch}_{\rho,\nu}(\texttt{q})$
in a series in $\texttt{q}$, coincides with the integers $N^+_a(\texttt{d}|\,{\cal S})$ with $-{\cal S}\le a <0$
and $\texttt{d}=|a|,\,|a|+1,\,|a|+2,\ldots\ $.
\medskip

Finally, we mention that the states $|\psi_{p,s}^{(S^z)}(\pmb{w})\rangle\in{\cal W}_{\rho,\nu}$
appearing in the scaling limit of the Bethe states, see eq.\eqref{asmnnb3298as}, have an important
interpretation. They are the simultaneous eigenstates of the
family of commuting operators known as the quantum AKNS integrable structure
\cite{Ablowitz:1974ry,Fateev:2005kx}. The function $\re^{\frac{\ri}{2}\delta}$ \eqref{asnbbvb32vbAAAAv} entering into the
quantization condition coincides with the eigenvalue of a certain
so-called reflection operator \cite{Zamolodchikov:1995aa} computed on 
$|\psi_{p,s}^{(S^z)}(\pmb{w})\rangle$, see ref.\cite{Kotousov:2019nvt} for details.

\section{Partition function in the scaling limit\label{sec5}}
In the case of the lattice model with (quasi-)periodic Boundary Conditions (BCs) imposed,
it was proposed in the work
\cite{Ikhlef:2011ay} and then verified numerically  in ref.\cite{Bazhanov:2019xvyA} that the
partition function appearing in the scaling limit of the lattice system, $Z^{\rm (scl)}$,  coincides with twice the partition function
for the 2D Euclidean black hole CFT. The latter was constructed in 
refs.\,\cite{Maldacena:2000kv,Hanany:2002ev}  by computing  
a functional integral with the worldsheet being taken to be a torus. 
The results presented in the previous section allow one to easily compute $Z^{\rm (scl)}$ for the 
staggered six-vertex model subject to $U_q\big(\mathfrak{sl}(2)\big)$ invariant open BCs.
One may expect $\frac{1}{2}\,Z^{({\rm scl})}$ to  coincide with the partition function for the 2D Euclidean black hole CFT
on the open segment $x\in(0,R)$, with certain conditions imposed on the fields at $x=0,R$. 

\medskip

Consider the lattice partition function 
\be
Z_L^{({\rm lattice})}={\rm Tr}_{\mathscr{V}_{2L}}\big[\re^{-M\,\mathbb{H}}\big]\,,
\ee
where the Hamiltonian  $\mathbb{H}$ is given by \eqref{askj8923} with 
$q=\re^{\frac{\ri\pi}{n+2}}$ and $n\ge 0$, while the trace is taken
over the $2^{2L}$ dimensional space of states: 
$\mathscr{V}_{2L}=\mathbb{C}^2_1\otimes\mathbb{C}^2_2\otimes\ldots \otimes \mathbb{C}^2_{2L}$. 
Keeping the ratio 
\be
\tau=\frac{v_{\rm F} M}{2L}
\ee
fixed as $L\to\infty$, one finds that the large $L$ behaviour  of the lattice partition function is given by
\be
Z_L^{({\rm lattice})}\asymp \re^{-MLe_{\infty}-Mf_{\infty}}\,Z^{({\rm scl})}\,.
\ee
Here $Z^{({\rm scl})}$ takes the form of a trace  over the space of states ${\cal H}$ appearing in the scaling limit
of the lattice model:
\be\label{kas89jhq3}
Z^{({\rm scl})}={\rm Tr}_{\cal H}\Big(\texttt{q}^{\hat{H}_{\rm CFT}}\Big)\qquad{\rm with}\qquad \texttt{q}=\re^{-2\pi\tau}\,.
\ee
It involves the `CFT Hamiltonian' $\hat{H}_{\rm CFT}$ which when restricted to the finite dimensional spaces 
${\cal H}^{(\rm{cont},\texttt{d})}_{p,s}$  
or ${\cal H}^{(\rm{disc},\texttt{d})}_{p,s}$ appearing in the decomposition of ${\cal H}$ 
coincides with the identity operator multiplied by the factor
\be\label{asmdnbb2n1oac}
E_{\rm CFT}=\frac{p^2}{n+2}+\frac{s^2}{n}
-\frac{1}{12}+\texttt{d}\ .
\ee
Notice that the asymptotic formula for the energy \eqref{asaaaaaakj90312} can be re-written as the formal relation
\be\label{asoi9832121}
\hat{H}_{\rm CFT}=\slim_{L\to\infty}\ 
\frac{L}{\pi v_{\rm F}}\,\Big(\mathbb{H}- L\,e_\infty-f_\infty\Big)\ .
\ee

\medskip

In subsection \ref{sec41} the space of states ${\cal H}$ was expressed as a direct sum of the continuous spectrum
${\cal H}^{({\rm cont})}$ and the discrete one ${\cal H}^{({\rm disc})}$, see formula \eqref{mn3287sdjh}.
The contribution of the states to the trace in eq.\,\eqref{kas89jhq3}
for each of these spaces will be denoted as $Z^{({\rm cont})}$ and $Z^{({\rm disc})}$,
respectively, so that
\be
Z^{({\rm scl})}=Z^{({\rm cont})}+Z^{({\rm disc})}\,,
\ee
where 
\be
Z^{({\rm disc})}={\rm Tr}_{{\cal H}^{({\rm disc})}}\,\Big(\texttt{q}^{\hat{H}_{\rm CFT}}\Big)\ ,\qquad\qquad
Z^{({\rm cont})}={\rm Tr}_{{\cal H}^{({\rm cont})}}\,\Big(\texttt{q}^{\hat{H}_{\rm CFT}}\Big)\ .
\ee
Let's first focus on the computation of $Z^{({\rm disc})}$.
The space ${\cal H}^{({\rm disc})}$ is made up of the components 
${\cal H}_{{\cal S},S^z}^{({\rm disc})}\cong{\cal H}_{{\cal S},{\cal S}}^{({\rm disc})}$, which admit the
decomposition \eqref{samnnb123AAA} into finite dimensional spaces.
Introduce the notation:
\be
\chi_{a,{\cal S}}(\texttt{q})=
\texttt{q}^{-\frac{(\frac{n}{2}-{\cal S}-a)^2}{n}+\frac{p^2}{n+2}-\frac{1}{12}}\ \bigg(\prod_{j=1}^\infty (1-\texttt{q}^j)^{-2}\bigg)\,\sum_{m=0}^\infty (-1)^m\,
\big(1-\texttt{q}^{(1+ m)(2{\cal S}+1)}\big)\,\texttt{q}^{ma+\frac{m(m+1)}{2}}\ ,
\ee
where, aside from the prefactor, the function $\chi_{a,{\cal S}}(\texttt{q})$ coincides with the generating function for 
the dimensions of the level subspaces ${\cal H}_{{\cal S},{\cal S}}^{({\rm disc,\texttt{d}})}$, see eqs.\,\eqref{ask98aaa32jd}
and \eqref{askjb32bvbvsd}. Then, the contribution of the discrete spectrum to the partition function reads as:
\be\label{ask893jkds}
Z^{({\rm disc})}=\sum_{{\cal S}\ge 0}(2{\cal S}+1)\bigg(\chi_{-{\cal S},{\cal S}}(\texttt{q})+2\!\!\!\!\!\sum_{a\in\mathbb{Z}\atop
0<a+{\cal S}<\frac{n}{2}}\!\!\!\!\!
\chi_{a,{\cal S}}(\texttt{q})\bigg)\ .
\ee
Each term in the sum over ${\cal S}$ has multiplicity
$(2{\cal S}+1)$ as a result of the $U_q\big(\mathfrak{sl}(2)\big)$ 
symmetry of the lattice model. Also, for every state with given $s=s_a$
\eqref{kjasnb1221sa} there exists another
one with $s=-s_a$ which yields the same contribution to the partition function,
except for the case when $s=\pm\frac{\ri n}{2}$, where they are identified as the same state.
This explains why the functions $\chi_{a,{\cal S}}(\texttt{q})$ come with a factor of two except 
the one with  $a=-{\cal S}$ (recall that 
$s_a=\ri(\tfrac{n}{2}-{\cal S}-a)$ 
and hence  $s_a=\frac{\ri n}{2}$ for $a=-{\cal S}$).
\medskip

The contribution of the continuous spectrum to the partition function is given by
\be\label{9832jdsa1}
Z^{({\rm cont})}=\sum_{{\cal S}\ge 0}(2{\cal S}+1)\,\int_{-\infty}^{\infty}\rd s\ 
\sum_{\texttt{d}\ge 0}\rho^{(\texttt{d})}_p(s)\,\texttt{q}^{\frac{s^2}{n}+\frac{p^2}{n+2}-\frac{1}{12}+\texttt{d}}\, .
\ee
Here $\rho^{(\texttt{d})}_p(s)$ is the density of states defined in formulae \eqref{kasnb12bdaaaa} and \eqref{mn2398jsc}, while
recall that
$2p=2{\cal S}+1-(n+2)$.
Notice that $Z^{({\rm cont})}$ becomes singular as $L\to\infty$:
\be
Z^{({\rm cont})}=Z^{({\rm sing})}+O(1)\, ,
\ee
where the singular part goes
as $\log(L)$ and reads as
\be
     Z^{({\rm sing})}= \sqrt{\frac{n}{2 \tau}}\ \frac{\log\big(2^{\frac{n+2}{n}}\,L/L_0\big)}
{\pi\,\texttt{q}^{\frac{1}{24}}\prod_{m=1}^\infty(1-\texttt{q}^m)}\ 
\sum^{\infty}_{\mathcal{S}=0} \,\, (2\mathcal{S}+1)\ \texttt{q}^{-\frac{1}{24}+\frac{p^2}{n+2}}\ 
\frac{1-\texttt{q}^{2\mathcal{S}+1}}{\prod_{m=1}^\infty (1-\texttt{q}^m)}\, .
\ee
The factor out the front of the sum is easily recognized to be the partition function of a boson
taking values in the segment $\sim\log(L)$ with Neumann BCs imposed at the endpoints of the
field at $x=0,R$. As for the remaining term,
\be\label{idsugewhbjnkas}
\sum^{\infty}_{\mathcal{S}=0} \,\, (2\mathcal{S}+1)\ \texttt{q}^{-\frac{1}{24}+\frac{p^2}{n+2}}\ 
\frac{1-\texttt{q}^{2\mathcal{S}+1}}{\prod_{m=1}^\infty (1-\texttt{q}^m)}\,,
\ee
in all likelihood it corresponds to a boundary state which is a superposition of Ishibashi states 
associated with a degenerate representation of the Virasoro algebra with generic central charge $c$ 
(see ref.\,\cite{Gaberdiel:2001zq} for the $c=1$ case). Note that \eqref{idsugewhbjnkas} also appears in the
scaling limit of the $XXZ$ spin $\frac{1}{2}$ 
chain with open $U_q\big(\mathfrak{sl}(2)\big)$ invariant BCs imposed \cite{PS1}.
\medskip

Formulae \eqref{ask893jkds}  and \eqref{9832jdsa1} do not seem to correspond to   the published results
in the literature concerning branes in the 2D Euclidean black hole CFT, in particular, ref.\,\cite{Ribault:2003ss}.
As such,  a separate investigation is required in order to establish a possible relation between the partition function
$Z^{({\rm scl})}=Z^{({\rm cont})}+Z^{({\rm disc})}$ and that of the black hole CFTs in the presence of boundaries.

\section{Discussion}
In this work the universal behaviour of the staggered six-vertex model with $U_q\big(\mathfrak{sl}(2)\big)$
invariant boundary conditions imposed was considered. We focused on the so-called self-dual case in the
critical regime \eqref{aksj8923j}. The problem was reformulated in the Hamiltonian picture, where
a central r\^{o}le belongs to $\mathbb{H}$ \eqref{askj8923}, which commutes with the
 transfer-matrix of the vertex model. The study of the $1/L$ corrections to the $L\to\infty$
behaviour of the energy of  $\mathbb{H}$ for the low energy states allowed us to 
 extract the spectrum of scaling dimensions of the statistical system. Our treatment, involving the use
of  novel numerical and analytic techniques, 
represents an advance on the type of analysis of the scaling limit of integrable lattice systems
with open boundary conditions that typically exists in the literature.
\medskip

The numerical construction of the RG trajectories was achieved via the method 
of the $Q$ operator. A key r\^{o}le was played by the
formula \eqref{asnb3bv} for the matrix elements of $\mathbb{Q}(\zeta)$, 
 valid for a one parameter family of open boundary conditions  in the sector $S^z=0$.
This is an original  result of our work.
The advantage of \eqref{asnb3bv},
as opposed to the expressions for $\mathbb{Q}(\zeta)$ appearing in the literature 
\cite{Tsuboi:2020uoh,Frassek:2015mra,Baseilhac:2017hoz,VlaarWeston}, is that it contains no infinite sums;
works literally for any (generic) complex values of $q$ and the boundary parameter $\epsilon$; and can
be programmed efficiently on the computer.
 We believe that, in view of the important applications
of the $Q$ operator, it may be worthwhile to 
extend \eqref{asnb3bv} to the other sectors of the Hilbert space with $S^z\ne 0$
and the case of more general open boundary conditions.
\medskip

The powerful analytic technique, which was crucial to our investigation, is the ODE/IQFT 
approach to the study of the scaling limit of integrable, critical lattice systems. For the
staggered six-vertex model in the regime \eqref{aksj8923j} with quasi-periodic boundary conditions imposed,
 it was developed in the works \cite{Bazhanov:2019xvy,Bazhanov:2019xvyA}.
We found that it was  applicable  to the 
 case of $U_q\big(\mathfrak{sl}(2)\big)$
invariant open boundary conditions as well. This points to the versatility of the  ODE/IQFT approach,
where the analysis for one set of  boundary conditions can be readily
carried over to another.
\medskip

As was already observed in 
 refs.\,\cite{Robertson:2020imc,Frahm:2021ohj}, the set of scaling dimensions of the statistical system
possesses a continuous
component, labeled by the quantum
number  $s=\slim_{L\to\infty} b(L)\in\mathbb{R}$ with $b(L)$ from eq.\,\eqref{ska823jhdsAAA} \cite{Frahm:2021ohj}. 
One of the  results of this work is the explicit formula for the density of states $\rho^{(\texttt{d})}(s)$ \eqref{kasnb12bdaaaa},\,\eqref{mn2398jsc},
which characterizes the continuous spectrum.  In addition, we studied the RG trajectories $\{|\Psi_L\rangle\}$, 
where $s$
becomes a pure imaginary number in the scaling limit. Building on the analysis of 
\cite{Frahm:2021ohj},  the discrete set   $\Sigma\equiv \Sigma^+\cup\Sigma^-$ \eqref{328daaa7ssa}
of all admissible 
 values of pure imaginary $s$ was found. We also determined
the dimension of the linear span of states
occuring in the scaling limit of $|\Psi_L\rangle$ with given $s\in\Sigma$
and conformal dimensions $\Delta$ \eqref{sa8721hdsb}.
\medskip

Our work includes a full characterization of the linear space ${\cal  H}$ appearing in the scaling limit of the
 space of low energy states 
 of the lattice system. To describe it,   ${\cal H}$ was decomposed  into a direct sum of the `continuous spectrum'
${\cal H}^{({\rm cont})}$ and the `discrete spectrum' ${\cal H}^{({\rm disc})}$. The
former, when expressed in terms of    finite dimensional spaces,
involves a direct integral over $s$, see eqs.\,\eqref{samnnb123},\,\eqref{sakj9832jh},\,\eqref{kasnb12bvad},
while the latter 
contains  a direct sum \eqref{samnnb123AAA}. We explained how the graded linear spaces
$\oplus_{\texttt{d}=0}^\infty {\cal H}_{p,s}^{({\rm cont},\texttt{d})}$ and 
$\oplus_{\texttt{d}=0}^\infty {\cal H}_{p,s}^{({\rm disc},\texttt{d})}$ are irreps. of the ${\cal W}_\infty$ algebra --
the algebra of extended conformal symmetry of the model.
\medskip

Perhaps the most interesting question is the relation between the scaling limit of the lattice system
and the 2D Euclidean/Lorentzian black hole CFTs \cite{Elitzur:1991cb,Mandal:1991tz,Witten:1991yr}. 
We believe that the formula for the partition
function $Z^{({\rm scl})}$ provided in sec.\,\ref{sec5} may be of help. Unfortunately, it does not seem
to correspond to known results in the literature on branes in the 2D black hole CFTs. It is likely
that progress in this direction would require a separate and detailed study.

\section*{Acknowledgments}
\noindent
The authors are grateful to M. Flohr and V. Schomerus for their interest in the work and for valuable discussions.
In addition, we would like to 
thank S. Ribault for including an enlightening discussion about  
the CFT interpretation of our results in his referee report. 
GK would also like to thank V. V. Bazhanov, S. L. Lukyanov and J. Teschner for useful and stimulating interactions.
\medskip

\noindent
The research of the authors is supported by the Deutsche Forschungsgemeinschaft (DFG) under grant No.
Fr 737/9-2. Part of the numerical work has been performed on the LUH computer cluster, which is funded
by the Leibniz Universit\"{a}t Hannover, the Lower Saxony Ministry of Science and Culture and the DFG.

\appendix

\section{The asymptotic coefficients \texorpdfstring{$\mathfrak{C}_{p,s}^{(\pm)}$}{CC} \label{AppB}}
Here we provide a closed form expression for the coefficients $\mathfrak{C}_{p,s}^{(\pm)}(\pmb{w})$
that were obtained in the work \cite{Kotousov:2019nvt}. Among other things, they enter into the quantization condition \eqref{kajs8712jhds1}.
\medskip

One has
\be
\mathfrak{C}^{(\pm)}_{p,s}
(\pmb{w})=\mathfrak{C}_{p,s}^{(0,\pm)}\,\check{\mathfrak{C}}^{(\pm)}_{p,s}(\pmb{w})\,,
\ee
where 
\be
\mathfrak{C}_{p,s}^{(0,\pm)}= 
\sqrt{\frac{2\pi}{n+2}}\ \ \ 2^{-p\pm\frac{\ri(n+2)s}{n}}\ 
(n+2)^{-\frac{2p}{n+2}}\ 
\frac{\Gamma(1+2p)}{\Gamma(1+\frac{2p}{n+2})\,\Gamma(\frac{1}{2}+p\pm\ri s)}\ ,
\ee
while $\check{\mathfrak{C}}^{(\pm)}$ are one for $\texttt{d}=0$. In the general case,
they are given by the determinant of a $\texttt{d}\times \texttt{d}$ matrix:
 \bea\label{hsaysayt}
\check{\mathfrak{C}}^{(\pm)}_{p,s}(\pmb{w})=\frac{(\mp 1)^\texttt{d}\ \det\big(w_a^{b-1}\,{ U}^{(\pm)}_a(b)\big)}
{\prod_{a=1}^\texttt{d} w_a\ \prod_{b>a}(w_b-w_a)\ \prod_{a=1}^\texttt{d}\big(2p+2a-1\pm 2 \ri s\big)}
\eea
with
\bea\label{aoisisaiaskjj}
{ U}^{(\pm)}_a(D)
&=&  (D-1)^2-
 \bigg( 2p+2+n\mp 2w_a+
\sum_{b\not= a}^\texttt{d}\frac{4 w_a}{w_a-w_b}\bigg)\ (D-1)\nonumber\\
&+&\tfrac{1}{2}\, n^2+\big(p+\tfrac{3}{2}\big)\ n\mp (n+1+2p+2\ri s)\ w_a+2 p+1
\\[0.2cm]
&+&\bigg(\sum_{b\not=a}^\texttt{d}\frac{2w_a}{w_a-w_b}\,\bigg)^2
+\, \big(
2 \,(2p+1+n\mp 2w_a)-n \,\big)\, \sum_{b\not=a}^\texttt{d}\frac{w_a}{w_a-w_b}\ .\nonumber
\eea

%%% FIRST OPTION
% Write your entries here directly, following the example below, including:
% Author(s), Title, Journal Ref. with year in parentheses at the end, followed by the DOI number.

%\begin{thebibliography}{99}
%\bibitem{1931_Bethe_ZP_71} H. A. Bethe, {\it Zur Theorie der Metalle. i. Eigenwerte und Eigenfunktionen der linearen Atomkette}, Zeit. f{\"u}r Phys. {\bf 71}, 205 (1931), \doi{10.1007\%2FBF01341708}.
%\bibitem{arXiv:1108.2700} P. Ginsparg, {\it It was twenty years ago today... }, \url{http://arxiv.org/abs/1108.2700}.
%\end{thebibliography}

\end{document}